%% file: North-South-Track-GRB-Paper-2016.tex
\documentclass[twocolumn]{aastex6}
\usepackage{graphicx}
\usepackage{amsmath}
\usepackage{hyperref}
\usepackage{natbib}
\usepackage{units}
\usepackage{hyperref}
\usepackage{bm}
\usepackage{lmodern}
\usepackage{color}
\usepackage{aas_macros}

\newcommand{\Bcal}{\mathcal{B}}
\newcommand{\Lcal}{\mathcal{L}}
\newcommand{\Scal}{\mathcal{S}}
\newcommand{\Tcal}{\mathcal{T}}

\newcommand{\unsim}{\mathord{\sim}}

\newcommand{\tabref}[1]{Table~\ref{tab:#1}}
\newcommand{\figref}[1]{Figure~\ref{fig:#1}}

\newcommand{\secref}[1]{Section~\ref{sec:#1}}

\begin{document}

\input{authors}

%\title{An All-Sky Search for Muon Neutrinos Coincident with
%Observed Gamma Ray Bursts in IceCube}
\title{Extending the search for muon neutrinos coincident with gamma-ray bursts \\ 
in IceCube data}
\slugcomment{Submitted to Astrophysical Journal}
%\slugcomment{To appear in Astrophys. J., v. 0.1}
\shortauthors{M.~G.~Aartsen et al.}
%\shorttitle{An all-sky muon neutrino search from GRBs with IceCube}
\shorttitle{Extending the search for muon neutrinos from GRBs with IceCube}

%\date{\today}

%\email{\myemail}

\begin{abstract}

  We present an all-sky search for muon neutrinos produced during the prompt
  $\gamma$-ray emission of 1172 gamma-ray bursts (GRBs) with the IceCube
  Neutrino Observatory. The detection of these neutrinos would constitute
  evidence for ultra-high energy cosmic ray (UHECR) production in GRBs, as
  interactions between accelerated protons and the prompt $\gamma$-ray field
  would yield charged pions, which decay to neutrinos. A previously reported
  search for muon neutrino tracks from Northern Hemisphere GRBs has been
  extended to include three additional years of IceCube data. A search for
  such tracks from Southern Hemisphere GRBs in five years of IceCube data has
  been introduced to enhance our sensitivity to the highest energy neutrinos. No
  significant correlation between neutrino events and observed GRBs is seen in
  the new data. Combining this result with previous muon neutrino track searches
  and a search for cascade signature events from all neutrino flavors, we
  obtain new constraints for single-zone fireball models of GRB neutrino and
  UHECR production.

\end{abstract}

\maketitle
\section{Introduction} 
\label{sec:intro}

The sources of ultra-high energy cosmic rays (UHECRs) with energies above
$\unit[10^{18}]{eV}$ remain unknown as intergalactic magnetic fields deflect
UHECRs while they propagate through the universe. In their source, interactions
of accelerated hadrons with matter or radiation are expected to produce high
energy neutral particles, namely photons and neutrinos. As these
particles are chargeless, they propagate through the universe undeflected,
meaning they can be associated with astrophysical sources to elucidate the
origin of UHECRs~\citep{Beatty.UHECRs.2009, KoteraOlinto.UHECRs.2011}.
Observations of $\gamma$-rays alone are insufficient to locate hadronic
accelerators as purely electromagnetic processes may also produce $\gamma$-rays
at these sources. Furthermore, $\gamma$-rays are not ideal messengers at the highest
energies because their propagation is hindered by interactions with interstellar
media or radiation, reducing their observable distance to the Local Group.
On the other hand, neutrinos only interact through the weak force and gravity,
allowing them to propagate from their source to Earth unimpeded. The detection
of high energy neutrinos from an astrophysical object would constitute
unambiguous evidence for hadronic acceleration, revealing UHECR
sources~\citep{Learned.HENuAstro.2000, HalzenHooper.HENu.2002,
AnchordoquiMontaruli.HENu.2010, Anchordoqui.HENu.2014}.

One possible class of sources for UHECRs are gamma-ray bursts (GRBs), which
release immense quantities of $\gamma$-ray radiation on time scales of
$\unit[10^{-3} - 10^3]{s}$. The predominant model for GRB phenomenology involves
the release of a relativistic fireball~\citep{Piran.GRBs.2004,
Meszaros.GRBs.2006, FoxMeszaros.GRBs.2006}---a plasma of electrons, photons, and
hadrons---that is triggered by a cataclysmic stellar collapse or binary system
merger. Shock waves present within the fireball are capable of accelerating
protons and electrons to very high energy through first-order
Fermi-acceleration~\citep{Krymskii.1977, Fermi.1949}.  As the relativistic
electrons are accelerated, they will radiate $\gamma$-rays that contribute to the
observable prompt $\gamma$-ray flux once radiative pressure expands the fireball
sufficiently so that the plasma becomes optically thin. Should protons in the
fireball be accelerated with comparable efficiency and abundance to electrons in
the fireball, the cosmological energy density of GRBs is sufficient to explain
the measured UHECR flux~\citep{Waxman.GRBUHECR1.1995, Vietri.GRBUHECR.1995}.
Additionally, these protons would interact with the ambient photon field to
produce high energy neutrinos primarily through the $\Delta^+$ resonance:

\begin{linenomath*}
\begin{equation}
  p + \gamma \rightarrow \Delta^+ \rightarrow n + \pi^+ \rightarrow n + e^+ +
  \nu_e + \bar{\nu}_{\mu} + \nu_{\mu}.
  \label{delta_res}
\end{equation}
\end{linenomath*}
These neutrinos, called \emph{prompt} neutrinos, would be observable in both
temporal and spatial coincidence with the prompt $\gamma$-ray emission of GRBs. 

The IceCube Neutrino Observatory~\citep{IC.FirstYear.2006, IC.Detector.2016} is currently the most
sensitive detector to astrophysical neutrinos.  An astrophysical
neutrino flux was discovered in neutrino interactions occurring within the detector
volume~\citep{IC.2yrHESE.2013, IC.3yrHESE.2014}, while observations of
$\nu_{\mu} + \bar{\nu}_{\mu}$ events from the Northern Hemisphere ($\delta >
-5^{\circ})$ later confirmed the discovery~\citep{IC.CosmicMuonNu.2015, IC.CosmicMuonNu.2016}.
IceCube has not yet observed a neutrino signal associated with GRBs \citep{IC.NorthTrackGRB.2010,
IC.NorthTrackGRB.2012, IC.NorthTrackGRB.2015, IC.AllskyCascadeGRB.2016}.
%, either in $\nu_{\mu} +
%\bar{\nu}_{\mu}$ induced \emph{track} events from the Northern Hemisphere with the
%partially completed detector~\citep{IC.NorthTrackGRB.2010,
%IC.NorthTrackGRB.2012} or the full detector~\citep{IC.NorthTrackGRB.2015}, or
%in \emph{cascade} signature events from all neutrino flavors over the entire
%sky~\citep{IC.AllskyCascadeGRB.2016}.  
These results are consistent with the
non-detection in multiple years of analysis by
AMANDA~\citep{AMANDA.AllskyCascadeGRB.2007, AMANDA.NorthTrackGRB.2008} and
ANTARES~\citep{ANTARES.TrackGRB.2013a, ANTARES.TrackGRB.2013b}. 
%The latest GRB search results reported by
%IceCube~\citep{IC.AllskyCascadeGRB.2016} combine possible contributions from
%individual GRBs into a single search by stacking the GRBs, allowing stringent
%limits to be placed on neutrino production in populations of GRBs. As these
%limits make a discovery from GRB stacking unlikely in the near-term, an
%individual treatment of GRBs is better suited to discovering possible neutrino
%emission from especially neutrino-bright GRBs.

This paper presents a continued search for prompt $\nu_{\mu} + \bar{\nu}_{\mu}$
neutrino track events from GRBs with IceCube~\citep{IC.NorthTrackGRB.2015} in
three additional years of data, but introduces two additional components to the
analysis: an analysis of each observed GRB individually, and an extension of the
Northern $\nu_{\mu} + \bar{\nu}_{\mu}$ track event search to the Southern Hemisphere
($\delta \leq -5^{\circ})$, where IceCube is most sensitive to the highest
energy neutrinos as Earth absorption attenuates this neutrino signal from the
Northern Hemisphere~\citep{Connolly.NuXSec.2011}.  Section~\ref{sec:grbs} describes the
prompt neutrino models tested in this analysis. Section~\ref{sec:icecube} then
reviews the IceCube detector and data acquisition. The neutrino candidate event
characterization and selections performed for the separate Northern Hemisphere track
and Southern Hemisphere track analyses are summarized in
Section~\ref{sec:reco_selection}. The unbinned maximized likelihood method for
discovery of a prompt neutrino signal from GRBs in both the stacked and per-GRB
contexts is then outlined in Section~\ref{sec:unbinned_likelihood}. Finally,
Section~\ref{sec:results} presents the results of our all-sky track analysis,
with Section~\ref{sec:conclusions} providing conclusions and an outlook for
future neutrino searches from GRBs with IceCube.

\section{GRB Prompt Neutrino Production}
\label{sec:grbs}

The search for neutrinos associated with GRBs in this paper considers only a
flux of neutrinos observable during the prompt stage of $\gamma$-ray emission,
and does not explicitly test precursor~\citep{Razzaque.GRBPrecursorNu.2003} or
afterglow~\citep{Waxman.GRBAfterglowNu.2000, Murase.GRBAfterglowNu.2006} models.
In the absence of an observed neutrino flux, we chose to place limits on two
classes of GRB models. The first class normalizes the expected neutrino flux to
that of the observed UHECR flux. The second class is more detailed and derives
an expected neutrino flux from the details of $\gamma$-ray emission for each of
the GRBs entering into the analysis.

The principal model describing the phenomenology of GRBs involves a beamed,
relativistic fireball of electrons, photons, and hadrons released from a black
hole central engine~\citep{Piran.GRBs.2004, Meszaros.GRBs.2006,
FoxMeszaros.GRBs.2006}. In the standard internal shock fireball model, particle
acceleration is achieved through first-order Fermi-acceleration at shock fronts
created in collisions of shells of plasmas moving at different speeds within the
fireball. The observed prompt $\gamma$-ray emission of GRBs can be produced
through synchrotron radiation and inverse-Compton scattering of accelerated
electrons or the decays of neutral pions to very energetic photons, while
accelerated hadrons within the fireball could possibly escape as UHECRs.

If one assumes that the highest energy cosmic rays are produced through the escape
of accelerated hadrons from GRBs, then a subsequent flux of neutrinos can be
calculated~\citep{Waxman.GRBNu.1997, Ahlers.GRBNu.2011}. The bulk Lorentz factor
$\Gamma$ of the fireball is largely unknown and thought to be in the range $100
\lesssim \Gamma \lesssim 1000$, though recent multi-wavelength observations of
several long GRBs have found values for $\Gamma$ as low as $\gtrsim
10$~\citep{Laskar.MultiWaveGRB.2015}. This value affects both the normalization
and spectral break energy of neutrinos produced in GRB fireballs. A benchmark
value of $\Gamma = 300$ is taken in the literature~\citep{Waxman.GRBNu.1997} to
calculate average neutrino spectra assuming average $\gamma$-ray emission
parameters, leading to neutrino spectra that are double broken power laws
peaking around $\unit[100]{TeV}$. We present flux limits within a range of
expected neutrino spectral break energies in Section~\ref{sec:results}.

Alternatively, one may assume that the predicted neutrino flux is related to the
observed $\gamma$-ray emission of individual GRBs~\citep{Hummer.RevGuetta.2012,
Zhang.AltFB.2013}, allowing direct limits to be placed on emission-generating
model parameters. Three representative models are tested, the internal shock
fireball model~\citep{Hummer.RevGuetta.2012, Zhang.AltFB.2013}, the photospheric
fireball model~\citep{Rees.GRBPhotosphere.2005, Murase.GRBPhotosphereNu.2008,
Zhang.ICMART.2011}, and the Internal Collision-induced Magnetic Reconnection and
Turbulence (ICMART) model~\citep{Zhang.ICMART.2011,Zhang.AltFB.2013}.
Phenomenologically, they primarily differ in neutrino production radius from the
GRB central engine, which scales the energy of the primary hadrons, and the
density and number of neutrino-producing interactions~\citep{Zhang.AltFB.2013}. 
%The photospheric model places the interaction radius at the photosphere
%($R_{\mathrm{ph}} \sim \unit[10^{10} - 10^{12}]{m}$), where the fireball
%transitions from optically thick to thin in $\gamma\gamma$ interactions. The
%internal shock model relates the collision radius to the variability time scale
%observed in $\gamma$-rays and $\Gamma$-factor of the fireball, and is thought to
%be approximately $R_{\mathrm{IS}} \sim \unit[10^{13} - 10^{14}]{m}$. The ICMART
%model pushes the interaction radius to $R_{\mathrm{ICMART}} \sim
%\unit[10^{15}]{m}$, where a shocks in the Poynting-flux-dominated disrupt the
%ordered magnetic fields, inducing particle acceleration through magnetic
%reconnection events. 
The neutrino spectra are calculated numerically using the Monte Carlo particle
interaction generator SOPHIA~\citep{Mucke.SOPHIA.2000}: protons are propagated
in an ambient $\gamma$-ray field with spectrum derived from that measured at
Earth, parameterized as a broken power law, and accounting for the redshift of the
GRB and the bulk Lorentz factor $\Gamma$ of the fireball. Neutrinos are produced
in a full simulation of possible $p\gamma$ interactions, accounting for
synchrotron losses of interaction products.
%accounting for synchrotron losses in inelastic $p\gamma$ interactions. 
We do not, however, consider Fermi acceleration of these
products---especially pions and muons---that might significantly enhance
neutrino production in GRBs~\citep{Klein.MuAcc.2013, Winter.MuAcc.2014}. The
neutrino fluence of a given GRB at Earth is then determined accounting for the
cosmological distance of the source and neutrino oscillations.  All models
considered here assume that proton acceleration occurs at a single location
where $\gamma$-rays are also produced and emitted.  In these cases, the
predicted prompt neutrino fluence will scale linearly with the proton content of
the fireball. When this acceleration location constraint is relaxed and a
dynamic GRB outflow is considered, the predicted prompt neutrino fluence is
significantly reduced, and well below the sensitivity of
IceCube~\citep{Bustamante.MultiZoneNuGRB.2015, Globus.MultiZoneNuGRB.2015}.

%TODO: reword some of this (too similar to Hellauer's paper)
Information about each GRB is gathered from the Gamma-ray Coordinates Network
(GCN)\footnote{\url{http://gcn.gsfc.nasa.gov}} and the Fermi GBM
database~\citep{Fermi.GBMCat2.2014, Fermi.GBMSpecCat2.2014}, and is compiled on
a publicly accessible
website\footnote{\url{http://grbweb.icecube.wisc.edu}}~\citep{Aguilar.GRBweb.2011}.
The temporal search window $T_{100}$ is defined by the interval between the
earliest reported start time $T_1$ and the latest reported stop time $T_2$ among
all observing satellites ($T_{100} = T_2 - T_1$), while the burst localization
is chosen from the most precise measurement reported. Similarly, the
$\gamma$-ray fluence, break energy, and observed redshift are used as inputs to
the neutrino emission calculation.  In some cases these values are not measured,
and in such cases we adopt conventions previously used by earlier
analyses~\citep{IC.NorthTrackGRB.2012, IC.NorthTrackGRB.2015,
IC.AllskyCascadeGRB.2016}.  We distinguish short GRBs,
$T_{100}$~$\leq$~$\unit[2]{s}$, from long GRBs, $T_{100}$~$>$~$\unit[2]{s}$.  In
both cases, if a measured photon fluence is unavailable, an average value of
$\unit[10^{-5}]{erg\,cm^{-2}}$ is used.  If the $\gamma$-ray break energy is
unmeasured, we assume a value of $\unit[200]{keV}$ for long GRBs and
$\unit[1000]{keV}$ for short GRBs. A redshift measurement is not available for
all GRBs; for these GRBs, we use values of 2.15 for long bursts and 0.5
for short bursts. If the redshift was measured, the isotropic luminosity can be
approximated from the redshift, photon fluence, and
$T_{100}$~\citep{Hummer.RevGuetta.2012}; otherwise, an average value of
$\unit[10^{52}]{erg\,s^{-1}}$ for long bursts and
$\unit[10^{51}]{erg\,s^{-1}}$ for short bursts is used. The variability time
scale is generally unknown, so we use $\unit[0.01]{s}$ for long GRBs and
$\unit[0.001]{s}$ for short GRBs, values that are consistent with current
assumptions in the literature~\citep{Baerwald.RevGuetta.2011,
Hummer.RevGuetta.2012, Zhang.AltFB.2013}.

\begin{figure}[t]
  \centering
  \includegraphics[width=0.48\textwidth]{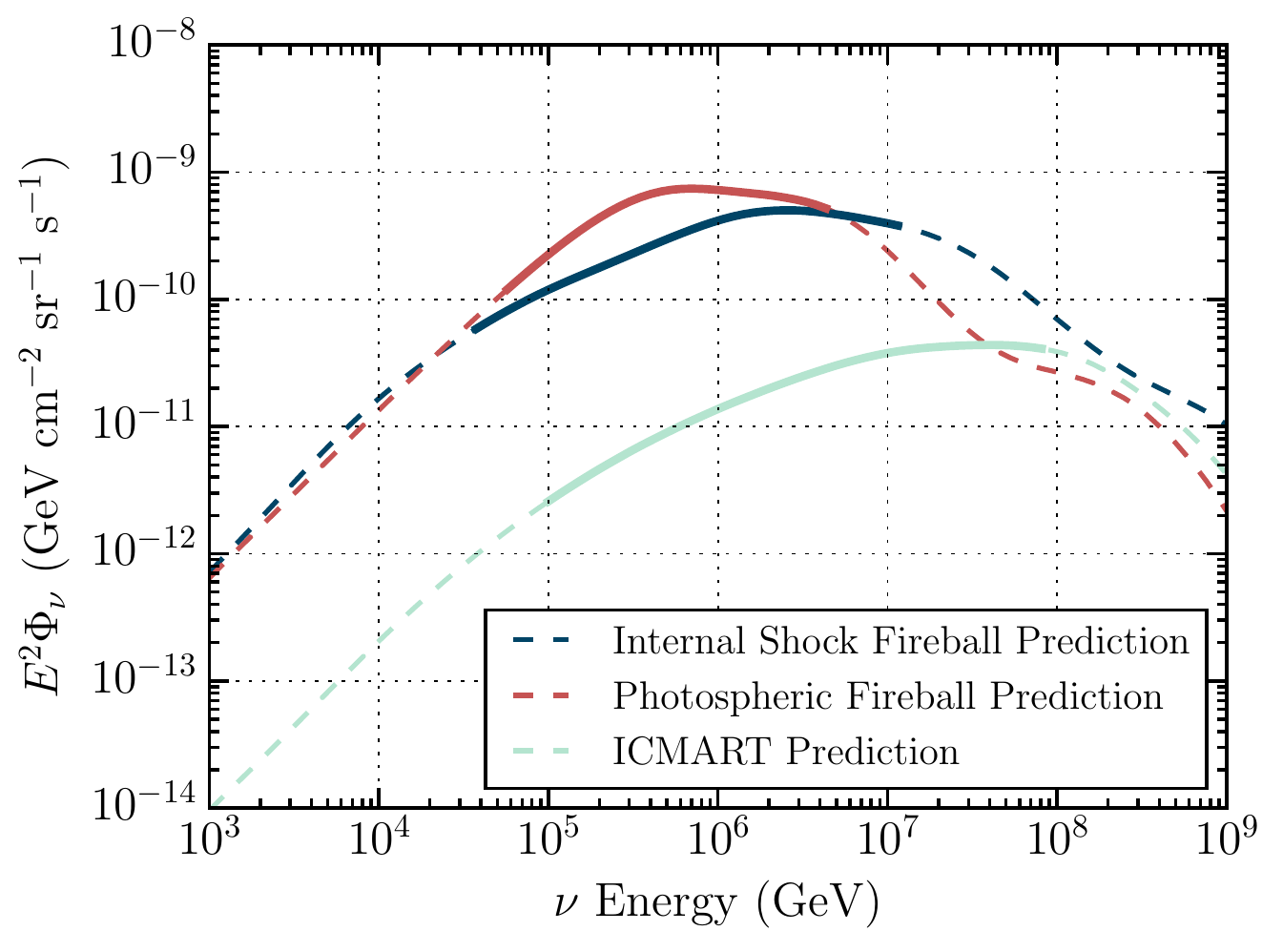}
  \caption{The predicted per-flavor quasi-diffuse flux of neutrinos from three
    numerical fireball models at benchmark parameters $f_p = 10$ and $\Gamma =
    300$ for the sample of 1172 GRBs analyzed here. The solid segments indicate
    the expected central 90\% energy containment interval of detected neutrinos.} 
  \label{fig:model_search_flux}
\end{figure}

Using benchmark model parameters of $\Gamma$~$=$~$300$ and a baryonic
loading---the ratio of fireball energy in protons to electrons---of
$f_{p}$~$=$~$10$, the expected model fluxes are shown in
\figref{model_search_flux}. We present the neutrino fluence calculations for the
analyzed sample of GRBs as a quasi-diffuse flux, assuming an average of 667
potentially observable GRBs per year distributed over the full sky, following
previous IceCube publications~\citep{IC.NorthTrackGRB.2012,
IC.NorthTrackGRB.2015, IC.AllskyCascadeGRB.2016}. Similar spectra can be
calculated for arbitrary values of $f_p$ and $\Gamma$.

\section{IceCube} 
\label{sec:icecube}

%Neutrino interactions can be detected in dense, optically clear media through
%the Cherenkov radiation of the relativistic charged secondary particles produced
%in these interactions. To detect astrophysical neutrinos produced in cosmic ray
%accelerator sources above the \emph{knee}, an estimated cubic kilometer of
%detector fiducial volume is required using ice or water as an interaction
%target~\citep{Halzen.km3nu.1993, Gaisser.km3nu.1995}. The IceCube Neutrino
%Observatory is the first detector to reach this detector volume benchmark, and
%has successfully detected these astrophysical neutrinos~\citep{IC.2yrHESE.2013,
%IC.3yrHESE.2014, IC.CosmicMuonNu.2015}.

IceCube consists of 5160 digital optical modules (DOMs) placed at depths from
\unit[1450]{m} to \unit[2450]{m} in the Antarctic ice shelf below the South
Pole~\citep{IC.FirstYear.2006, IC.DeepCore.2012}. Each DOM consists of a
photomultiplier tube (PMT) within a glass pressure sphere that detects the
Cherenkov radiation of neutrino interaction products~\citep{IC.PMT.2010}. The
photon signal measured by the PMT is digitized in the DOM and relayed to the
surface computing laboratory~\citep{IC.DAQ.2009}. The detector is arranged in a
hexagonal array of 86 vertical cables, each connected to 60 DOMs, called
\emph{strings}. The DOMs are placed on the strings at \unit[17]{m} intervals,
while the inter-string spacing is $\unsim \unit[125]{m}$. Above the in-ice
IceCube detector, 81 ice-tank pairs (each tank containing two DOMs) compose the
\emph{IceTop} surface array~\citep{IC.IceTop.2013}.
%, which has sensitivity to
%cosmic ray extensive air showers with energy $\gtrsim \unit[300]{TeV}$.
In this paper, the IceTop array is used to veto likely atmospheric muon events
from the sky above the detector whose air shower deposits Cherenkov radiation in
the surface tanks.

%The topology of neutrino interactions detected by IceCube are dependent on the
%interaction type. The charged current (CC) interactions of muon neutrinos yield
%a muon which can travel for many kilometers in the ice before decaying,
%appearing as well-localized \emph{tracks} of light in IceCube. Alternatively,
%the CC interactions of electron neutrinos and tau neutrinos, as well as neutral
%current (NC) interactions of all neutrino and anti-neutrino flavors, yield
%particle showers that, due to particle and photon scattering in the ice, appear
%as nearly-spherical \emph{cascades} within the detector. At sufficient energies,
%the interaction and decay vertices of tau neutrino CC interactions are expected
%to be separated by distances of $\unsim \unit[50]{m} \times (E_{\nu_{\tau}} /
%\unit[1]{PeV})$, yielding two distinguishable cascades called a
%\emph{double-bang} event. No double-bang events have yet been observed in
%IceCube, but should they be observed, due to low expected background rates, they
%would be an unambiguous channel of astrophysical neutrinos.

Muons produced by $\nu_{\mu}$ or $\bar{\nu}_{\mu}$ charged current (CC)
interactions---appearing as long \emph{tracks} of light within the IceCube
detector---are an especially convenient channel in searches for high energy
astrophysical neutrino sources. This is due to their long propagation length in the
ice, which increases the effective interaction volume of the detector and yields
sub-degree resolution in reconstructed muon direction from the long lever-arm
in the detector. Track events from astrophysical sources, however,
are difficult to disentangle from IceCube's primary backgrounds: atmospheric
muons and $\nu_{\mu} + \bar{\nu}_{\mu}$ produced in cosmic ray extensive air
showers.  Atmospheric muons produced in the sky above IceCube trigger the
detector at $\gtrsim \unit[2]{kHz}$, while atmospheric $\nu_{\mu} +
\bar{\nu}_{\mu}$ trigger the detector at $\unsim \unit[20]{mHz}$ over the full
sky.
%Muons produced by $\nu_{\mu} + \bar{\nu}_{\mu}$ charged current (CC)
%interactions are an especially convenient channel in searches for high energy
%neutrino production at astrophysical sources. These muons can travel for many
%kilometers in the ice before decaying, appearing as well-localized \emph{tracks}
%of light in IceCube. Due to the long ``lever-arm'' of light deposited in
%IceCube, reconstruction of muon direction with sub-degree angular resolution is
%possible. Reconstructions to nearly-spherical cascade events---produced in
%neutral current (NC) neutrino interactions and $\nu_e$ and $\nu_{\tau}$ CC
%interactions---conversely typically yield directions with angular resolutions of
%$\unsim 10^{\circ}$. Additionally, the long propagation length of muons in the
%ice greatly increases IceCube's effective volume to these events. Searches for
%track events from astrophysical sources, however, are difficult to disentangle
%from IceCube's primary backgrounds: atmospheric muons produced in cosmic ray
%extensive air showers in the Southern Hemisphere sky which trigger the detector
%at rates $\gtrsim \unit[2]{kHz}$, and CC interactions of atmospheric $\nu_{\mu}
%+ \bar{\nu}_{\mu}$ produced in the same cosmic ray air showers, which trigger
%the detector at $\unsim \unit[20]{mHz}$ over the full sky.
In recent searches for track events coincident with GRBs, only the Northern
Hemisphere sky was analyzed to effectively remove the atmospheric muon
background, with the irreducible atmospheric neutrino background remaining.
In those analyses, declination $\delta > -5^{\circ}$ is chosen to
define the Northern Hemisphere, as above these declinations no atmospheric
muons can reach the detector. Such searches have a
diminished sensitivity to the highest energy neutrino events compared to the
Southern Hemisphere sky ($\delta \leq -5^{\circ}$), as Earth absorption becomes
relevant for neutrinos with energy $\gtrsim \unit[1]{PeV}$, though this is
partially ameliorated by an increased sensitivity to signal near the analysis
horizon. The continued non-detection of a neutrino signal from GRBs compels the
extension of these track searches to the Southern Hemisphere.

IceCube data are searched for track events consistent with $\nu_{\mu}$ and
$\bar{\nu}_{\mu}$ CC interactions. The previously published Northern Hemisphere
GRB track analysis included data from three years of the partially completed
IceCube detector and one year of the completed
detector~\citep{IC.NorthTrackGRB.2015} and included 506~GRBs.  This has been
extended to include three additional years of IceCube data between May 2012 and
May 2015, during which 508~GRBs occurred in the Northern Hemisphere during good
detector operation.  The newly introduced Southern Hemisphere GRB track analysis
has been applied to IceCube data between May 2010 and May 2015, during which
IceCube operated for a year in a 79-string configuration and four years of the
full IceCube detector. During good detector operation, 664~GRBs occurred in the
Southern Hemisphere sky. In total, we searched for neutrino emission from
663~new GRBs, while 509~GRBs that were included in the all-sky cascade GRB
analysis~\citep{IC.AllskyCascadeGRB.2016} were analyzed for the first time in
the $\nu_{\mu} + \bar{\nu}_{\mu}$ track channel. 

\section{Event Reconstruction and Selection} 
\label{sec:reco_selection}

High quality track events were selected in IceCube data by the topology of the
particle light deposition. Likelihood-based reconstructions fit a track
hypothesis to the timing and position of PMT photoelectron pulses of a given
event, accounting for photon scattering and absorption in the ice, to obtain the
muon's direction~\citep{AMANDA.MuonReco.2004, IC.PS.2014}. Expected light yield
probability distribution functions (PDFs) are either defined analytically or
from spline fits to simulated light yields from cascades or minimally ionizing 
muons~\citep{Whitehorn.Photospline.2013, IC.PS.2014}. By parameterizing the
behavior of the likelihood space as a function of muon direction near the
reconstructed best-fit direction, the angular uncertainty in these
reconstructions can be estimated, allowing further selection of high-quality track
events. For use in the significance calculation in this analysis, the
reconstructed angular uncertainty is determined using the Cramer--Rao lower
bound~\citep{Cramer.CRbound.1945, Rao.CRbound.1945} on the covariance of angular
direction measures from the inverse of the likelihood Fisher information matrix.
Finally, the muon energy as it reaches the IceCube detector, as well as its
individual stochastic losses, can also be reconstructed through similar
likelihood-based fits to the measured light deposition.
These algorithms yield reconstructed muon energies with a resolution of $\unsim
30\%$, and a resolution of the total deposited energy in the detector along the
muon track of $10 - 15\%$~\citep{IC.MuonE.2014}.

The track samples in the Northern Hemisphere and Southern Hemisphere are
obtained separately as the primary background in each is fundamentally
different. In the Northern Hemisphere event sample, most events in low-level
data are actually atmospheric muon events from the Southern Hemisphere that are
misreconstructed to have an origin from the Northern Hemisphere. The event
selection primarily focuses on removing these poorly reconstructed
muons. In the Southern Hemisphere event sample, nearly all the events are
well-reconstructed atmospheric muons, which must be separated from the muon
signal produced in $\nu_{\mu} + \bar{\nu}_{\mu}$ CC interactions. In both cases,
the background is characterized using events that are more than $\pm
\unit[2]{hr}$ (termed \emph{off-time} data) from the prompt $\gamma$-ray
emission of any analyzed GRB, which avoids contamination of a possible GRB
neutrino signal.  The event selection was optimized to maximize the retention of
a Monte Carlo simulation of interactions of diffuse astrophysical $\nu_{\mu} +
\bar{\nu}_{\mu}$ neutrinos with an $E^{-2}$ spectrum, giving the selection
sensitivity to the wide range of neutrino production models of
\secref{grbs}. 

The Northern Hemisphere track sample was obtained following the same selection
as the Northern Hemisphere GRB track analysis presented by
\cite{IC.NorthTrackGRB.2015}. The background in this portion of the sky
is dominated by atmospheric muons with misreconstructed direction. A
selection optimized to well-reconstructed $\nu_{\mu} + \bar{\nu}_{\mu}$ signal
can efficiently remove most of this background. The parameters that effectively
distinguish these events have been described in previously published IceCube
point source~\citep{IC.PS.2011} and GRB~\citep{IC.NorthTrackGRB.2015} searches.
These parameters are used in a boosted decision tree (BDT)
forest~\citep{Freund.AdaBoost.1997}, a multivariate machine learning algorithm
that scores the effective signal-ness of a candidate event, to robustly separate
background off-time data from signal simulation. By removing events below a
certain BDT value, the most background-like events are
eliminated~\citep{IC.PS.2014, IC.AllskyCascadeGRB.2016}, arriving at the final
Northern Hemisphere event sample with a rate of $\unsim
\unit[6]{mHz}$. This data sample is dominated ($\unsim80\%$) by the irreducible
atmospheric $\nu_{\mu} + \bar{\nu}_{\mu}$ background from the Northern
Hemisphere, with the remainder of the background being composed of
misreconstructed atmospheric muons. The final BDT score cut was chosen such that
the discovery and limit-setting potential of the stacked unbinned likelihood
analysis (\secref{unbinned_likelihood}) was approximately optimized for a signal
with an $E^{-2}$ spectrum produced by all analyzed GRBs, following
\cite{IC.AllskyCascadeGRB.2016}. 
%The cut is then constrained by the discovery potential of the newly-introduced
%per-GRB analysis to an $E^{-2}$ spectrum produced by a single, randomly selected GRB
%in the sample; this signal simulates the situation where a visible neutrino fluence is
%produced in a small subset of the analyzed GRBs.
The cut may be further adjusted to coincide with the optimal discovery potential
of the newly-introduced per-GRB analysis for an $E^{-2}$ spectrum produced by a
single, randomly selected GRB in the sample, simulating a detectable neutrino
fluence produced in a small subset of the analyzed GRBs.  Though an $E^{-2}$
spectrum was used in the optimization, the GRB sensitivity was robust against
variation of signal spectra in both Northern and Southern Hemisphere analyses.
The $E^{-2}$ spectrum was therefore chosen for generality and consistency with
previous IceCube GRB analyses.

The Southern Hemisphere track selection was modeled on recent IceCube
point source analyses~\citep{IC.PS.2014, IC.PS.2017} in this portion of
the sky. This selection first removes the bulk of the low-energy atmospheric
muon background through cuts on single parameters. Machine learning is then employed
to reduce the atmospheric muon background further, especially the
high-multiplicity bundles of muons produced concurrently in high-energy cosmic
ray air showers and traversing the IceCube detector together.  These muon bundles
deposit large amounts of light in the detector, and are difficult to distinguish
from single high-energy muons by reconstructions of event energy.
%The atmospheric muons that dominate the Southern sky background have a much
%softer spectrum than the expected neutrino spectra from GRBs. Most of the
%selection therefore focusing on eliminating muons with energy $\lesssim
%\unit[100]{TeV}$ through energy proxy cuts as a function of muon inclination.
%Additionally, the most vertically inclined air showers are expected to trigger
%the IceTop detector. Thus, events with sufficient IceTop activity are removed
%from the analysis. At this point, the background primarily consists of high
%energy bundles of multiple muons that are produced concurrently in air showers
%and traverse the detector together. These muon bundles can be distinguished from
%single high energy muons through parameterization of their light deposition
%characteristics: 1)~the stochastic energy losses of multiple muons in the bundle
%yield a smoother apparent average energy deposition than the stochastic energy
%losses of a single muon, and 2)~the overlapping Cherenkov emission of the muons
%within a bundle will yield more early light compared to the Cherenkov emission
%of a single muon. 
A BDT forest is trained to distinguish atmospheric muons and muon bundles from
the well-reconstructed muons of simulated interactions of $\nu_{\mu}
+ \bar{\nu}_{\mu}$ with an $E^{-2}$ spectrum. Parameters supplied to the BDT
forest are those used in the point source
analyses~\citep{IC.PS.2014, IC.PS.2017}, as well as a number of new
parameters: 1)~an azimuthal measure of the event to regularize artificially
preferred directions in background events due to the IceCube detector geometry,
2)~the distance of the estimated neutrino interaction vertex from the detector
edge measured along the reconstructed track direction to select lower energy
neutrino starting tracks~\citep{IC.MESE.2016}, and 3)~the reconstructed muon
energy and zenith.  These additional parameters result in a more efficient
signal selection than the Southern Hemisphere point source selection at all
neutrino energies. A cut on the per-event BDT score yields the final event
sample (optimized under the same procedure as the Northern Hemisphere event
selection), with a background data rate of $\unit[2-3]{mHz}$ that is still
dominated by atmospheric muons ($\unsim 4\%$ atmospheric neutrinos). 

\begin{figure}[t]
  \centering
  \includegraphics[width=0.48\textwidth]{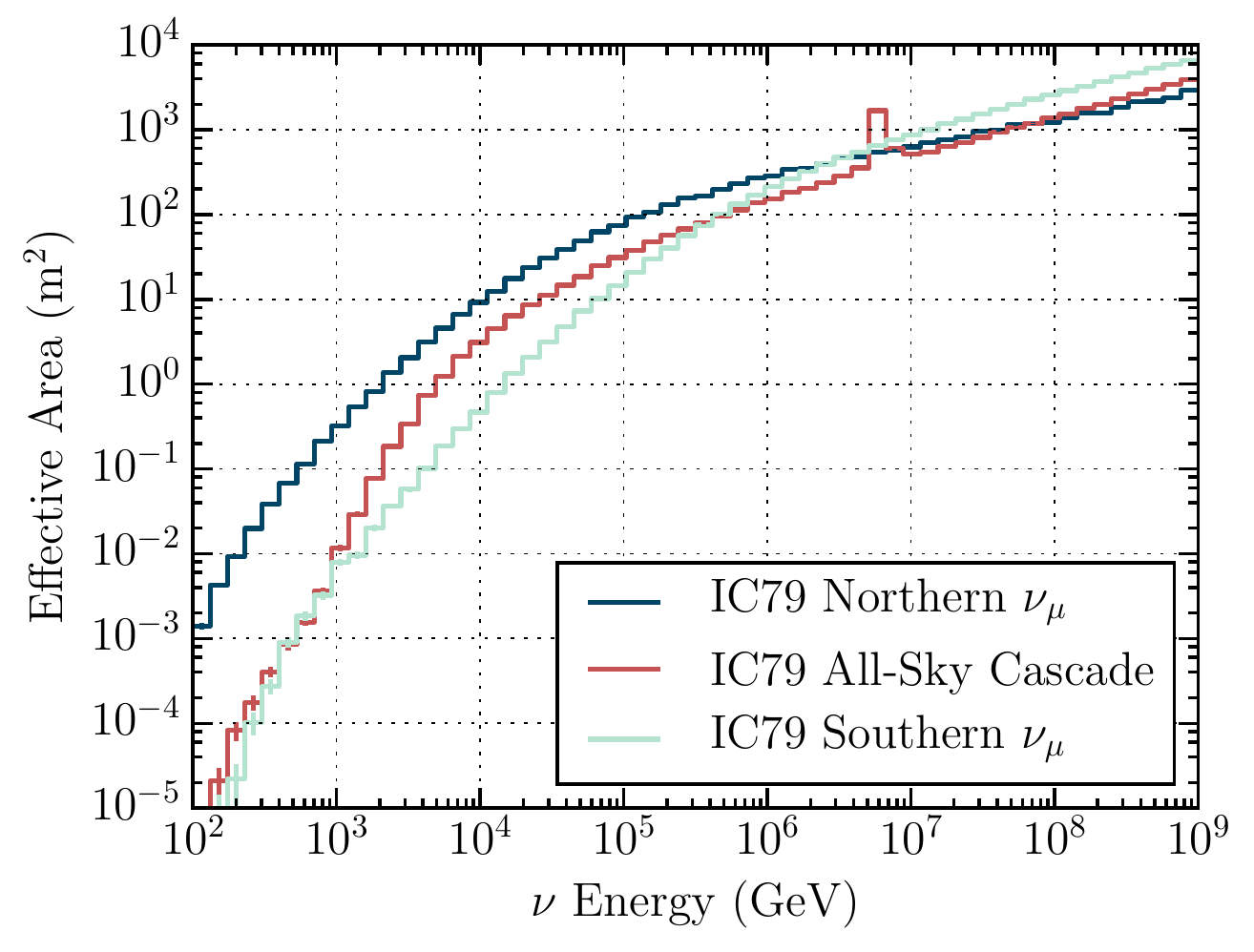}
  \caption{Effective areas, scaled to all-sky, of the Northern and Southern Hemisphere $\nu_{\mu}$
  track analyses compared to that of the all-sky cascade analysis for the
  79-string IceCube detector configuration.} 
  \label{fig:effective_area}
\end{figure}

The final expected signal event rate in this event selection can be
determined from Monte Carlo simulation of $\nu_{\mu} + \bar{\nu}_{\mu}$
interactions through

\begin{equation}
  \dot{N}_{\mathrm{signal}} = \int_{\Omega} d\Omega' \int dE_{\nu} A_{\mathrm{eff}}
  (E_{\nu}, \Omega') \times \Phi_{\nu} (E_{\nu}, \Omega'),
  \label{sig_rate}
\end{equation}
where $A_{\mathrm{eff}} (E_{\nu}, \Omega')$ is the effective area of neutrino
interaction for an event selection, $\Phi_{\nu} (E_{\nu}, \Omega')$ is the
signal neutrino flux, and the integral is performed over the analysis solid
angle $\Omega$ and neutrino energy $E_{\nu}$ range.  The effective areas, scaled to all-sky, of the
Northern and Southern Hemisphere track selections are shown in
\figref{effective_area}, compared to the all-sky cascade selection of
\cite{IC.AllskyCascadeGRB.2016}. The Northern Hemisphere selection is
demonstrated to be most sensitive to neutrinos with energy $\lesssim
\unit[1]{PeV}$, while the effective area of the Southern Hemisphere selection
displays the enhanced sensitivity of this channel to neutrinos above a few PeV.
The resonant scattering of $\bar{\nu}_e$ with electrons in ice at
$\unit[6.3]{PeV}$~\citep{Glashow.Resonance.1960} is seen in the all-sky cascade
effective area, and is yet to be observed by IceCube.

\section{Unbinned Likelihood Analysis} 
\label{sec:unbinned_likelihood}

Given an ensemble of neutrino events and a set of GRBs, a statistical test is
required to distinguish an observation of prompt neutrinos from expected
backgrounds. For a sample of $N$ events coincident with GRBs, we calculate the
significance of the coincidences by an unbinned likelihood with observed number of
signal events $n_s$ of the form:

\begin{equation}
\Lcal \left( n_{s} | n_{b}, \{ \bm{x}_{i} \}\right) = P_{N}\,\prod_{i=1}^{N}
  \left[ p_{s} \Scal \left( \bm{x}_{i} \right)\, +\, p_{b} \Bcal \left(
  \bm{x}_{i} \right) \right],
\end{equation}
where $p_{s} = n_{s} / (n_{s} + n_{b})$, $p_{b} = n_{b} / (n_{s}+n_{b})$, and
$P_N$ is the Poisson probability of the observed event count $N$ given expected
signal and background event counts $n_s$ and $n_b$, respectively:

\begin{equation}
P_{N} = \frac{\left(n_{s}+n_{b}\right)^{N}\,e^{-\left(n_{s}+n_{b}\right)}}{N!}.
\end{equation}
The index $i$ runs over the neutrino candidate events, and $\mathcal{S}$ and
$\mathcal{B}$ respectively represent the combined signal and background
probability density functions (PDFs) for event characteristics $\bm{x}_i$.  Each
of the signal and background PDFs is defined with respect to the time and
direction relative to the GRBs, and with respect to event energy.  The final
test statistic is the logarithm of the likelihood, maximized with respect to
$n_s$ (maximized at $\hat{n}_s$) divided by the background-only likelihood
($n_{s}=0$), which simplifies to:

\begin{equation}
  \Tcal = \ln \left[ \frac{\Lcal \left( \hat{n}_s \right)}{\Lcal \left( n_s = 0
        \right)} \right] = -\hat{n}_{s} + \sum_{i=1}^{N} \ln \left[
        \frac{\hat{n}_{s} \Scal \left(
        \bm{x}_{i} \right)}{\left< n_{b} \right> \Bcal \left( \bm{x}_{i}
  \right)} + 1\right].
\end{equation}
The average expected number of background events can be determined from
off-time data, denoted as $\langle n_b \rangle$.

\begin{figure}[t]
  \centering
  \includegraphics[width=0.4\textwidth]{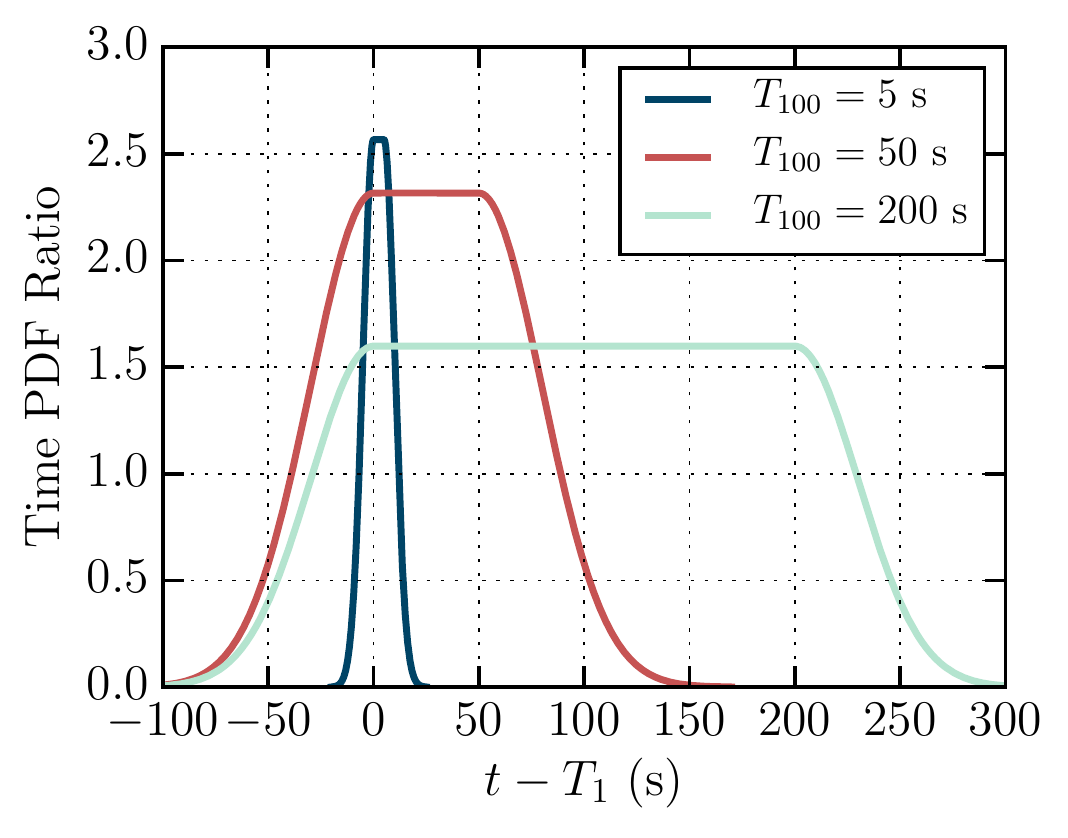}
  \caption{Signal-to-background PDF ratios for three GRB durations. The
    earliest reported start time $T_{1}$, and the latest reported stop time
    $T_{2}$, define the most inclusive GRB duration $T_{100}$.} 
  \label{fig:pdf_ratio_time}
\end{figure}

The time component of the signal and background PDFs, shown as a
signal-to-background PDF ratio in \figref{pdf_ratio_time}, is defined by the
$T_{100}$ of each burst.  The signal time PDF is constant during $T_{100}$,
with Gaussian tails before and after the GRB prompt phase. The
functional form of the Gaussian tails is chosen to have a smooth
transition on either side, and the Gaussian standard deviation $\sigma_T$ is
chosen to be the same as $T_{100}$, but limited to minimum and maximum
values of $\unit[2]{s}$ and $\unit[30]{s}$, respectively. For simplicity, the
signal time PDF is truncated after $\pm4\sigma$ in each of the Gaussian tails.
The background time PDF is constant in this search time window.

\begin{figure}[t]
  \centering
  \includegraphics[width=0.4\textwidth]{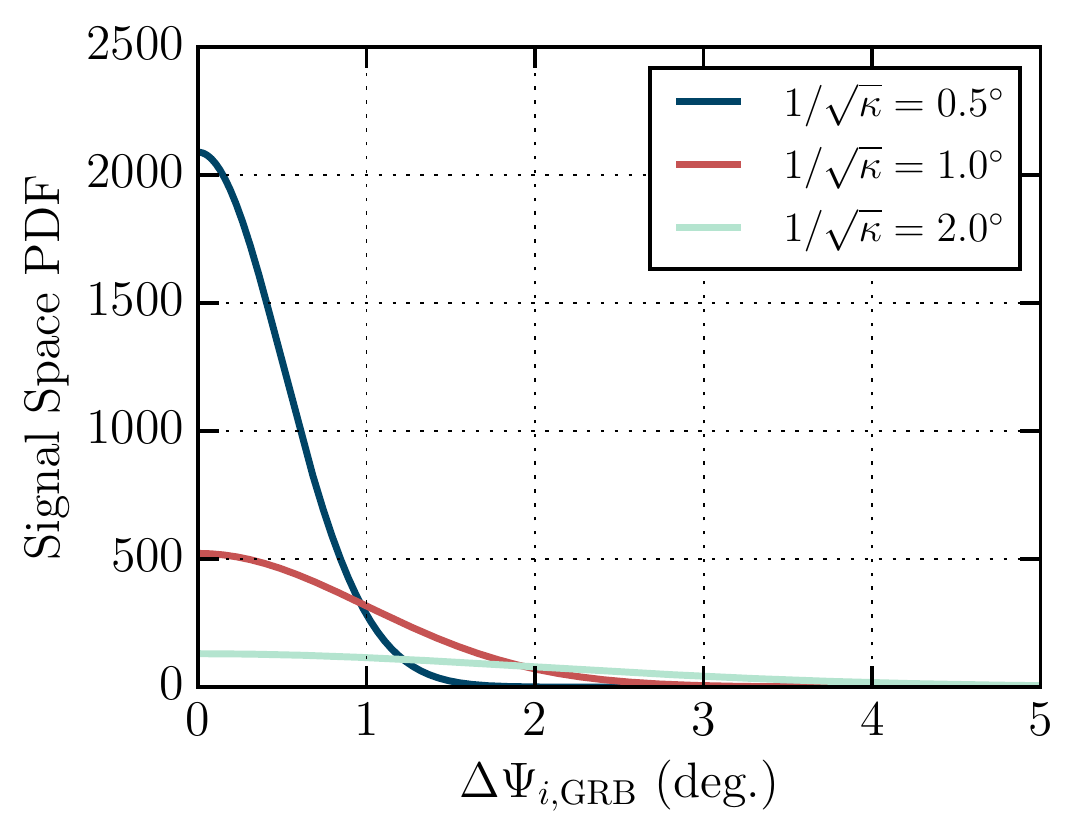}
  \caption{Kent distributions used for the signal space PDF for a number of
    concentration parameters $\kappa$.} 
  \label{fig:pdf_sig_space}
\end{figure}

\begin{figure*}[t!]
  \centering
  \includegraphics[width=0.45\textwidth]{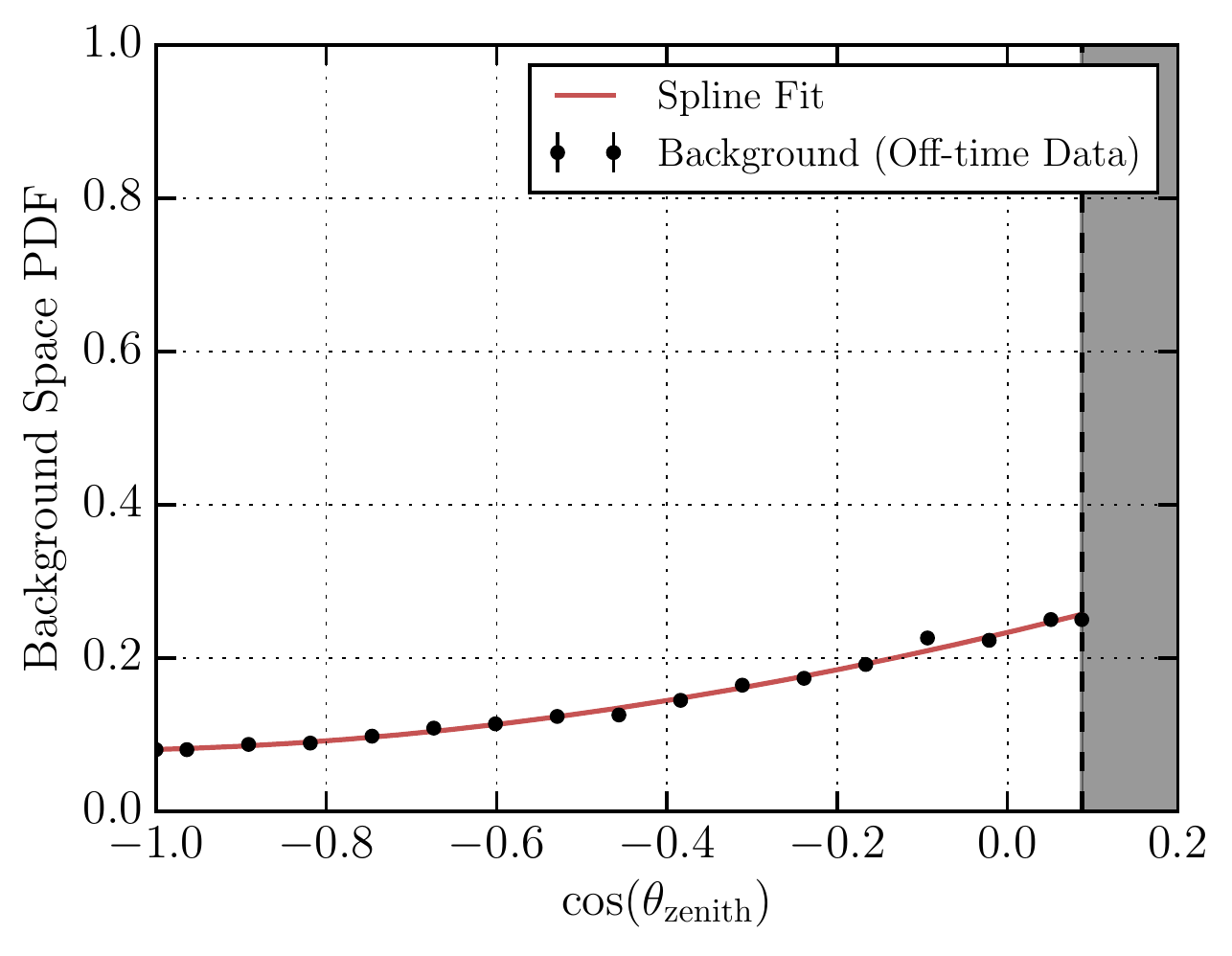}\hspace{10pt}
  \includegraphics[width=0.45\textwidth]{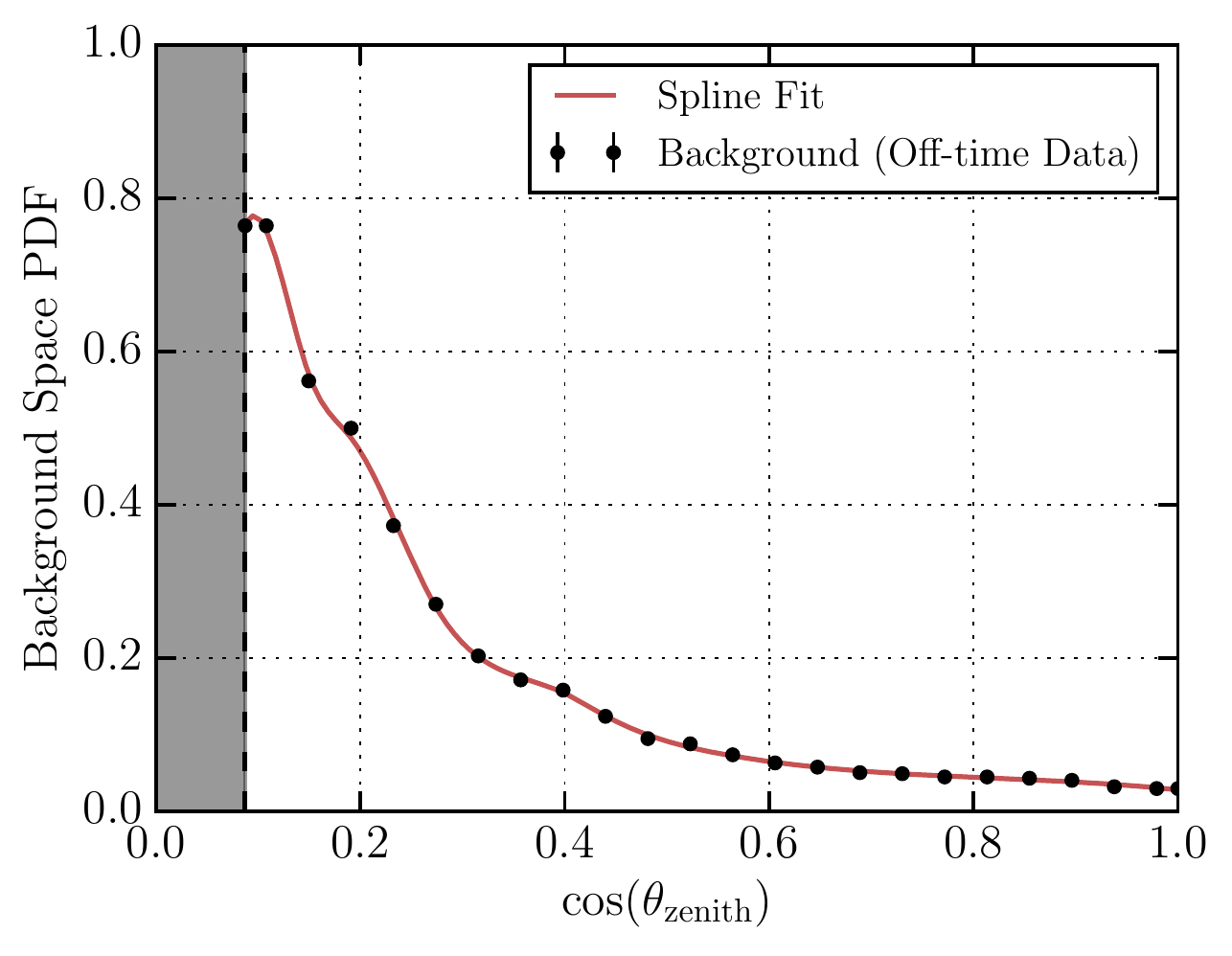}
  \caption{Background space PDF as a function of the cosine of the reconstructed
    event zenith angle for the Northern Hemisphere (left) and Southern Hemisphere
    (right) $\nu_{\mu}$ track analyses for binned off-time data (black points) and
    spline fit (red line). Each is normalized in its respective search solid
    angle. The analysis horizon is indicated by the dashed black line.} 
  \label{fig:pdf_bg_space}
\end{figure*}

Signal neutrinos from GRBs are expected to be spatially associated with the
observed GRB location. We define a PDF following the first-order
non-elliptical component of the Kent distribution~\citep{Kent.Dist.1982}:

\begin{equation}
  \Scal_{\mathrm{space}} (\vec{x}_i) = \frac{\kappa}{4\pi \sinh(\kappa)} e^{\kappa \cos
  (\Delta \Psi_{i,\mathrm{GRB}})}
  \label{kent_dist}
\end{equation}
where $\Delta \Psi_{i,\mathrm{GRB}}$ is the opening angle between the
reconstructed event direction and GRB location, and the concentration term
$\kappa$ is given by $\kappa = \left(\sigma_{i}^{2} +
\sigma_{\textrm{GRB}}^{2}\right)^{-1}$ in units of radians. The Kent
distribution is normalized on the unit sphere and is more appropriate than the
typical two dimensional Gaussian representation, especially for 
events with large uncertainties in the reconstructed direction.
The two dimensional Gaussian distribution is recovered for large concentration
parameters ($1 / \sqrt{\kappa} \lesssim 10^{\circ}$). Representative examples of
the Kent distribution with varying directional uncertainties are shown in
\figref{pdf_sig_space}. Data from the off-time sample are used to characterize
the background space PDF. Due to the azimuthal symmetry of the IceCube detector,
the background can be sufficiently described using only the zenith angle, with
PDF normalized over the solid angle of each analysis.  A spline is fit to a
histogram of background data in $\cos\left(\theta_{\mathrm{zenith}}\right)$. The
data histograms and spline fits for the Northern and Southern Hemisphere muon
neutrino searches are shown in \figref{pdf_bg_space}.

\begin{figure*}[t!]
  \centering
  \includegraphics[width=0.49\textwidth]{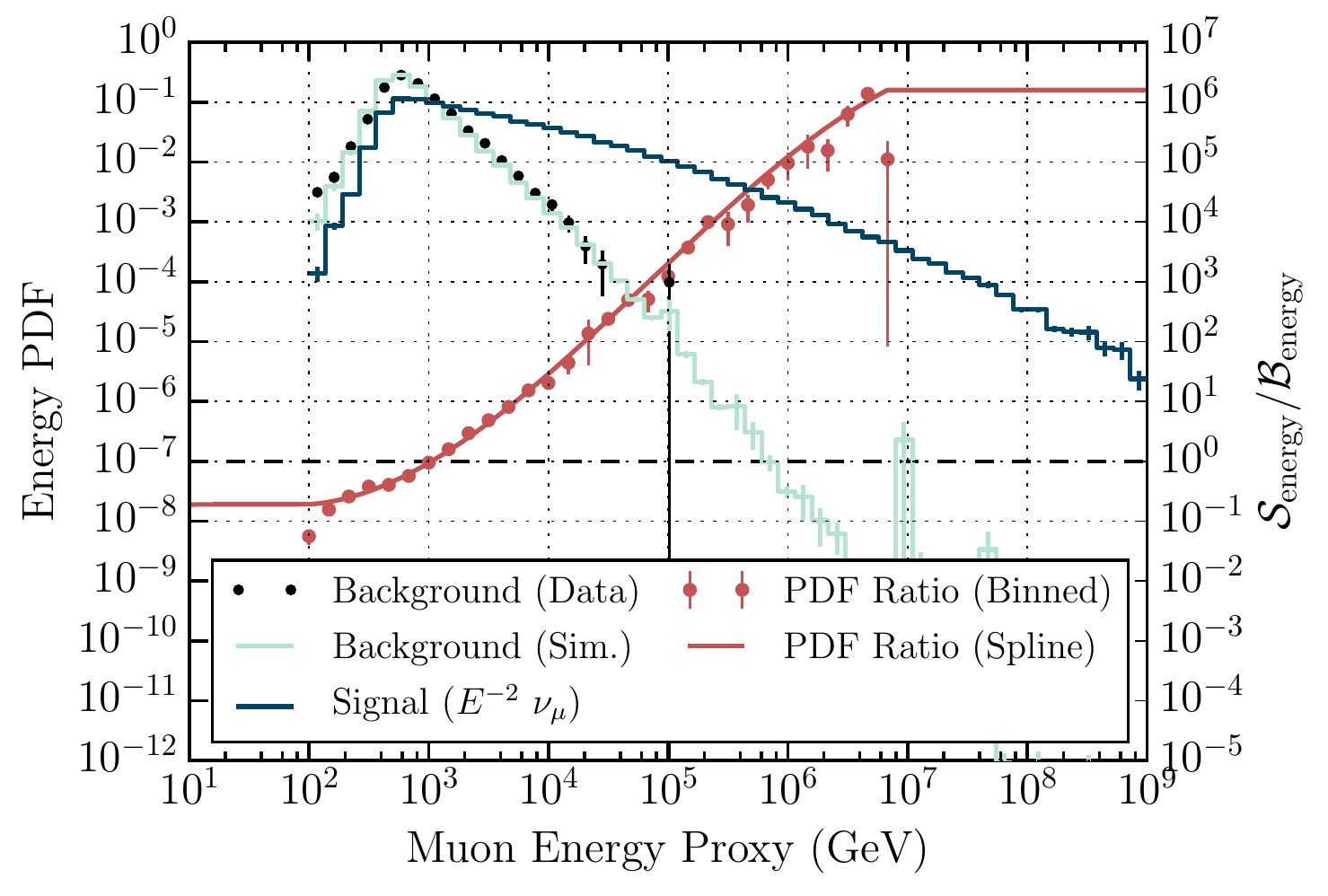}
  \includegraphics[width=0.49\textwidth]{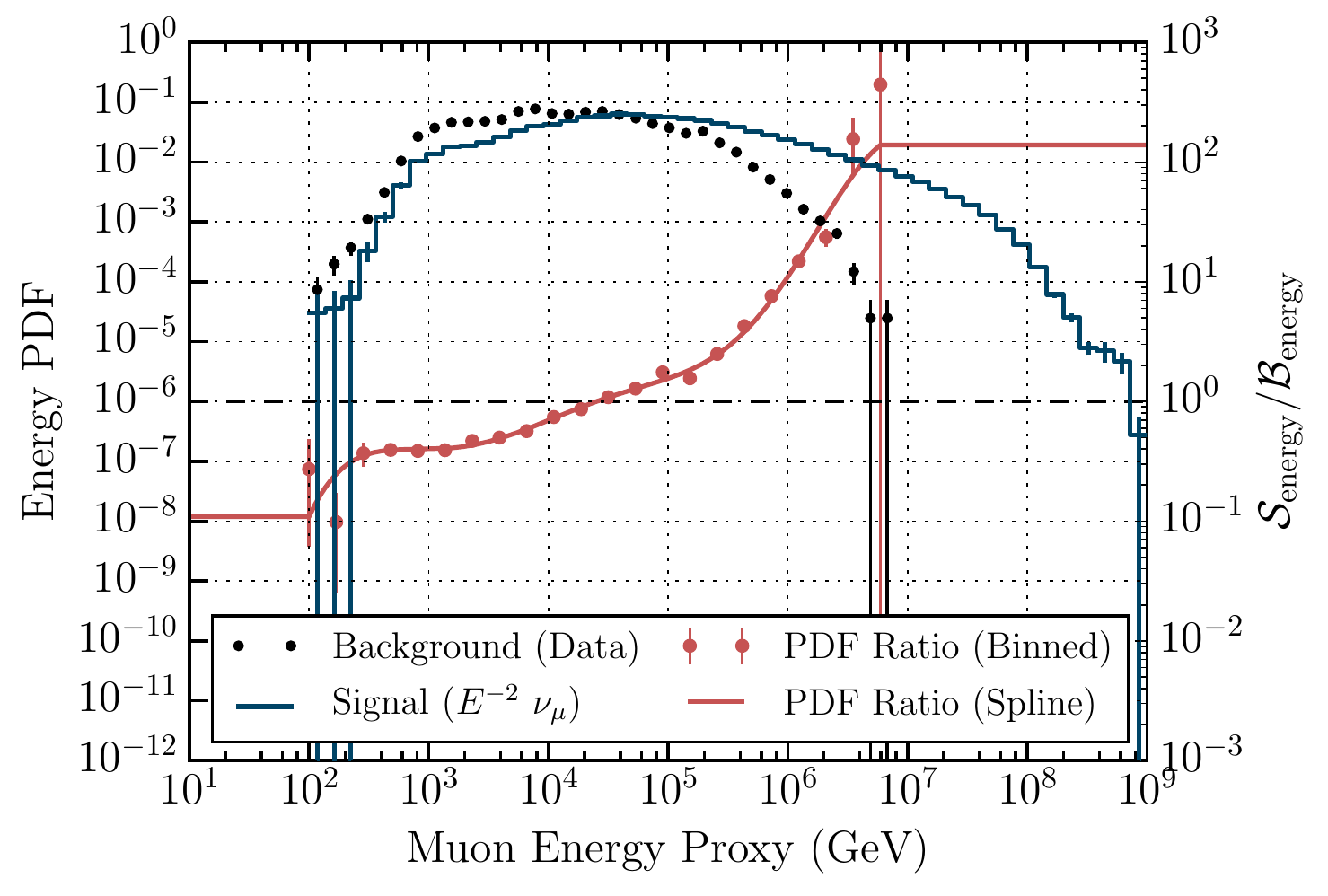}
  \caption{Energy PDFs and signal-to-background ratios for the Northern Hemisphere
    (left) and Southern Hemisphere (right) $\nu_{\mu}$ track analyses.  Left
    vertical axis: the reconstructed muon energy PDFs of background off-time
    data (black points) and $E^{-2}$ $\nu_{\mu}$ signal simulation (blue line);
    simulated background used for PDF extrapolation is provided in the Northern
    track analysis (green line). Right vertical axis: per-bin PDF ratios (red
    points) and spline fit (red line).}
  \label{fig:e_pdf_ratio}
\end{figure*}

One of the most powerful characteristics expected to distinguish GRB
neutrinos from atmospheric neutrinos is their respective energy spectra: prompt
neutrinos are expected to be produced at high energies where the steeply falling
atmospheric spectrum is significantly diminished.  The reconstructed energy of
muon tracks is used as a proxy for the incoming neutrino energy. Although the
muon energy proxy is only a lower bound, it scales with the neutrino energy and
is therefore still useful for distinguishing signal from background.  For
generality, the signal energy PDF is calculated using the reconstructed muon
energies of simulated neutrino events with an $E^{-2}$ spectrum. The background
energy PDF is taken directly from the off-time data sample's reconstructed
energy spectrum.  The Northern and Southern Hemisphere reconstructed energy
PDFs are shown in \figref{e_pdf_ratio} along with the binned PDF ratio
values as a function of reconstructed muon energy. The binned PDF ratio is fit
with a spline to generalize the ratio to arbitrary reconstructed energies.
At high and low energies where the distributions become sparsely populated, the
PDF ratio is conservatively limited to the value of the nearest bin with
sufficient statistics. In the Northern Hemisphere analysis, the background is
largely made up of atmospheric neutrinos. As such, the background energy PDF can
be artificially extended to very high energies by using simulated neutrino
events with an atmospheric spectrum~\citep{Honda.AtNuSpectrum.2007}.  The same
technique is not valid for the Southern Hemisphere analysis because the
background is composed of atmospheric muons. The simulation of cosmic ray air
shower events is significantly more computationally intensive, meaning the
statistics of the off-time data sample greatly exceeds our atmospheric muon
simulations at final cut level. Furthermore, the simulated atmospheric muon events
do not include a simulation of the IceTop detector, invalidating the comparison
of the simulation to the off-time data set, which includes an IceTop veto
selection. Thus, only off-time data is used to characterize the background
energy PDF in the Southern Hemisphere analysis.

%\begin{figure*}[t!]
%  \centering
%  \includegraphics[width=0.49\textwidth]{ic86ii_south_stacked_tsd_bdt_gt_0_250.pdf}
%  \includegraphics[width=0.49\textwidth]{ic86ii_south_max_pb_tsd_bdt_gt_0_250.pdf}
%  \caption{Test statistic distributions for the stacked (left) and $\max(\{
%    \Tcal_g \})$ (right) analysis methods for the IC86-2012 Southern Hemisphere
%    analysis. The background-only test statistic distribution (blue) is shown
%    relative to distributions in trials with both background and signal
%    ($E^{-2}$ spectrum) injection. Signal is injected under two different
%    hypotheses: a distributed signal where each GRB contributes an equal
%    neutrino fluence (red), and a signal which is concentrated in a single
%    random GRB within the sample (green). In both cases, the total detectable
%    neutrino fluence is $E^{2} dN/dE = \unit[0.5]{GeV\, cm^{-2}}$. The vertical
%    black lines indicate the median (solid), $3\sigma$ (dashed), and $5\sigma$
%    (dotted) background-only threshold test statistic values.}
%  \label{fig:ic86ii_south_tsd}
%\end{figure*}

In previous IceCube searches for prompt neutrinos from GRBs, the search was
performed by stacking all GRBs in each year and channel (i.e.\ Northern
Hemisphere track, Southern Hemisphere track, all-sky cascade). This method,
however, diminishes the significance of a concentrated neutrino signal from a
single GRB within the stacked sample, as the test statistic treats such events
equivalently to if they were distributed among all the GRBs in the sample. To
increase sensitivity of the analysis to a signal concentrated in individual
GRBs, we have adopted the new strategy of calculating a test statistic
$\Tcal_{g}$ for each GRB $g$. We then determine the GRB for which the maximal
$\Tcal_{g}$ value is obtained (called the $\max(\{ \Tcal_g \})$ method). This
approach improves the discovery potential for a signal from a single
neutrino-bright GRB and naturally moves into real-time style searches, as each
GRB would be treated individually upon detection. The $\max(\{ \Tcal_g \})$
method is preferred to the selection of the most significant per-GRB coincidence
(calculated relative to the background-only $\Tcal_g$ distribution for GRB $g$)
like that done in the IceCube point source searches, as it is less
computationally intensive and yields comparable signal discovery potential. The
trials-corrected significance of the final $\max(\{ \Tcal_g \})$ is determined
through a comparison with the background expectation of this test statistic
calculated over the entire analyzed Northern and Southern Hemisphere GRB sample.

The stacked test statistic was also still calculated for each year and channel
to possibly discover a weak neutrino signal distributed over multiple GRBs. This
allows the results presented in this paper to be combined with previous
results~\citep{IC.NorthTrackGRB.2012, IC.NorthTrackGRB.2015,
IC.AllskyCascadeGRB.2016}. A final stacked test statistic is calculated as a sum
over individual channel $c$ and year $s$ test statistics:

\begin{equation}
  \Tcal = \sum_{c,s} \Tcal_{c,s}.
\end{equation}
This combined test statistic is used to calculate limits on the GRB neutrino
models of \secref{grbs} as it is less sensitive to possible background
fluctuations than the per-GRB method. 

\begin{figure}[t]
  \centering
  \includegraphics[width=0.48\textwidth]{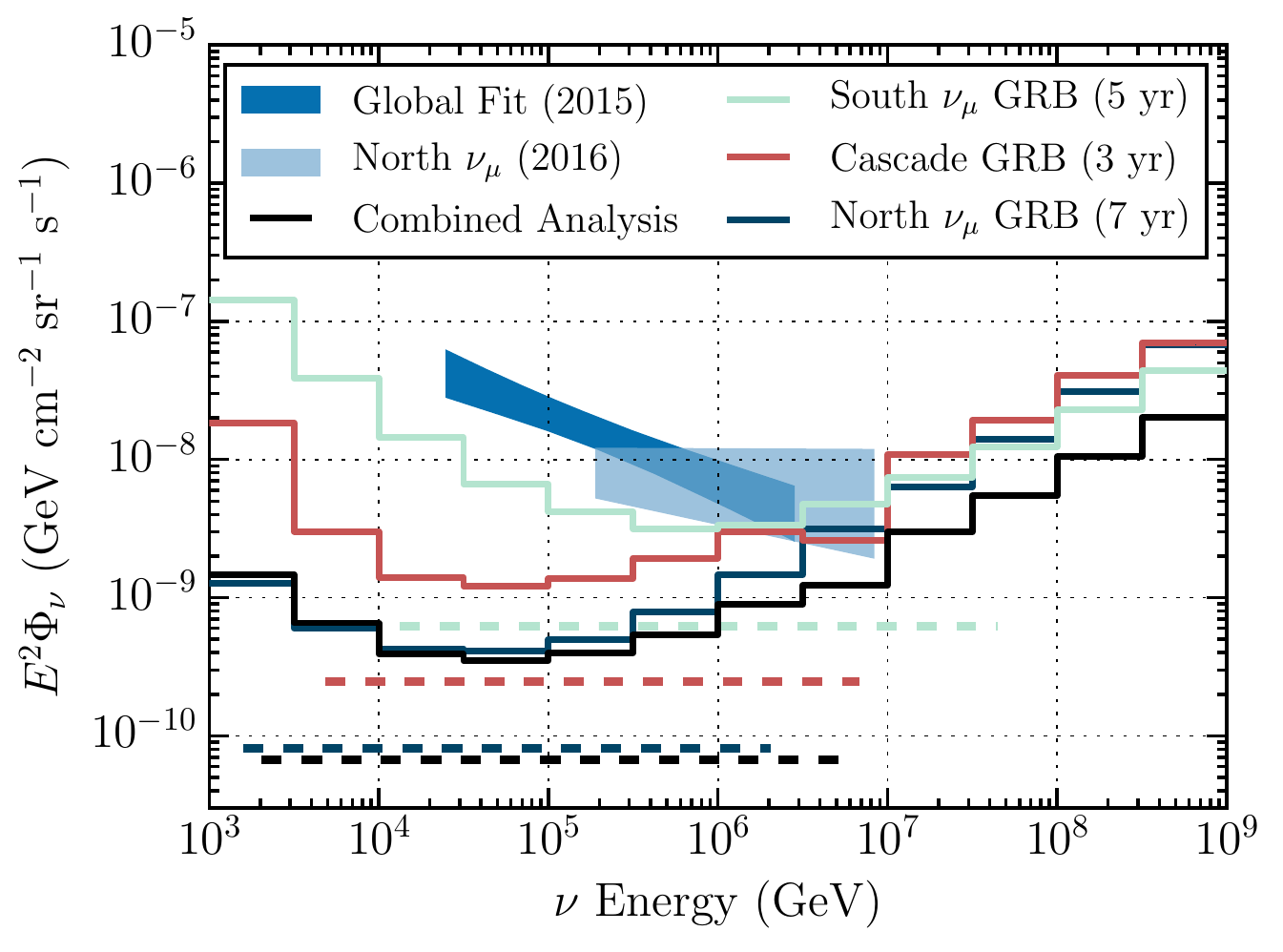}
  \caption{Differential median sensitivity of the Northern Hemisphere track,
    all-sky cascade~\citep{IC.AllskyCascadeGRB.2016}, and Southern Hemisphere
    track stacked GRB analyses to a per-flavor $E^{-2}$ $\nu$ quasi-diffuse flux
    in half-decadal $\nu$ energy bins, with the final combined analysis shown in
    the black line.  Integrated sensitivities are shown as dashed lines over
    the expected 90\% energy central interval in detected neutrinos for a given
    analysis. The IceCube measured 68\% CL astrophysical per-flavor neutrino flux
    band is given for reference from a global fit of IceCube
    analyses~\citep{IC.GlobalFit.2015} and a recent 6-year Northern Hemispheres
    $\nu_{\mu}$ track analysis (light blue, \cite{IC.CosmicMuonNu.2016}).} 
  \label{fig:combined_diff_sens}
\end{figure}

The background-only and background-plus-signal expectations of both stacked and
per-GRB analyses are determined from Monte Carlo pseudo-experiments following
the same methodology as described by \cite{IC.AllskyCascadeGRB.2016}. 
%Pseudo-experiments including neutrino signal
%simulation are then used to determine the discovery and limit-setting potential
%of a given analysis. 
%Test statistic distributions for both the stacked and $\max
%(\{\Tcal_g\})$ methods are shown in \figref{ic86ii_south_tsd} for the IC86-2012
%Southern Hemisphere analysis.  Here, the background-only test statistic
%distributions are compared to distributions with an $E^{-2}$ neutrino signal
%with a total fluence normalization of $E^{2} dN/dE = \unit[0.5]{GeV\, cm^{-2}}$
%included under two signal hypotheses: a distributed signal where each GRB
%contributes an equal neutrino fluence, and a signal which is concentrated in a
%single random GRB within the sample. As can be seen, the stacked method
%diminishes the significance of a large fraction of trials of the single
%neutrino-bright GRB hypothesis by stacking that signal with many other GRBs
%without a detected neutrino signal. Conversely, the $\max (\{\Tcal_g\})$
%method enhances the significance of signal under this source hypothesis, while
%reducing the significance of a distributed neutrino signal. Thus, both methods
%are employed given their sensitivity to different distributions of GRB neutrino
%production strengths.
The sensitivity, both differential and integrated, of the stacked method to a
per-flavor quasi-diffuse $E^{-2}$ neutrino spectrum is shown in
\figref{combined_diff_sens}. This sensitivity is calculated for each individual
search channel, as well as the final combined sensitivity. The Northern
Hemisphere track analysis (combining the results of \cite{IC.NorthTrackGRB.2015}
with this paper's extension to three additional years) is seen to be the most
sensitive neutrino detection channel. The all-sky cascade and Southern
Hemisphere track channels converge in sensitivity to the Northern Hemisphere
track within a factor of a few at energies $\gtrsim \unit[1]{PeV}$, while the
Southern Hemisphere track analysis is the most sensitive GRB analysis to date
for neutrinos $\gtrsim \unit[10]{PeV}$. Each individual channel has sufficient
sensitivity to detect a neutrino signal should the per-flavor quasi-diffuse GRB
neutrino flux be comparable in magnitude to the measured IceCube astrophysical
neutrino flux of $\unsim \unit[10^{-8}]{GeV\, cm^{-2}\, sr^{-1}\, s^{-1}}$.

\section{Results} 
\label{sec:results}

The final event sample was searched in coincidence with the 508
GRBs of the three-year Northern Hemisphere sample and the 664 GRBs of the
five-year Southern sample. Both per-GRB and stacked per-year and channel test
statistics were calculated to discover a neutrino signal from GRBs. The
results of the per-GRB analysis are presented for the Northern and Southern
Hemisphere analyses in Tables~\ref{tab:nt_results} and \ref{tab:st_results},
respectively.  Here, basic information about the GRBs and coincident events are
described, including their timing, angular uncertainty $\sigma$, angular separation
$\Delta \Psi$, the measured $\gamma$-ray fluence of the GRB, and the estimated
energy of the coincident event. The significance of the coincidences is
summarized in two ways. Event signal-to-background PDF ratio values used in the
test statistic calculation are provided to estimate relative event importance.
The significance of the per-GRB test statistic is then given as a p-value
calculated from that GRB's expected background-only test statistic distribution,
constituting that GRB's pre-trials p-value. In parentheses, the post-trials
p-value of this GRB coincidence is given, calculated relative to the combined
three-year Northern Hemisphere track and five-year Southern track analysis $\max
(\{\Tcal_g\})$ test statistic distribution expected from background, respectively.

The most significant coincidence (in both pre-trials and post-trials p-value)
was found in the Southern Hemisphere analysis coincident with GRB110207A, a
Swift-localized long GRB ($T_{100} = \unit[109.32]{s}$) observed at a
declination of $-10.8^{\circ}$. This event occurred during the $T_{100}$ of the
GRB and had a reconstructed direction within $1^{\circ}$ of the GRB, with a
moderate reconstructed muon energy of $E_{\mu} \gtrsim \unit[12]{TeV}$, yielding
a signal-to-background PDF ratio of $\Scal / \Bcal = 271.6$. The pre-trials
significance is $p = 3.5\times 10^{-4}$, making it the single most significant
coincidence with a GRB to date in any IceCube GRB neutrino search.  Although the
event was within $1^{\circ}$ of the GRB location, the angular uncertainty of
this event and GRB were $0.3^{\circ}$ and $0.01^{\circ}$, respectively.
Combined, these lead to a $\unsim 3\sigma$ offset in the signal space PDF,
reducing the significance of the coincidence.  Monte Carlo simulations and
reconstructions were performed of muons with similar energy, origin, and light
deposition topology to the measured event, establishing that the reconstructed
angular uncertainty of $0.3^{\circ}$ is consistent with the median angular
resolution of the simulated muons of $0.24^{\circ}$. Furthermore, a full
likelihood scan of a more detailed angular reconstruction, which accounts for
muon stochastic losses, was performed on this event to verify the quality of the
reconstructed direction~\citep{IC.MuonE.2014}. It was found that the two
reconstructions are consistent with each other, while the GRB110207A location is
$> 5\sigma$ from the advanced reconstructed direction, supporting that this
event is inconsistent with the GRB location. Additionally, the post-trials
significance of this event is $p = 0.535$, making it consistent with the
background-only hypothesis. Considered together, this event is concluded to be a
background coincidence event.

% TODO: Finish this paragraph
Two additional coincident events were observed in the Northern Hemisphere track
analysis that had event significances of $\Scal/\Bcal \gtrsim 100$: one event in
coincidence with GRB131202B, a Fermi-GBM localized long GRB ($T_{100} =
\unit[86.02]{s}$) at a declination of $21.3^{\circ}$, and one in coincidence with GRB150428B, a Swift localized
long GRB ($T_{100} = \unit[161.8]{s}$) at a declination of $4.1^{\circ}$. Both events occurred during the
$T_{100}$ of the GRBs, and had reconstructed deposited energies above $\unit[1]{TeV}$. 
Due to the short tracks these events produced, each had a relatively
large angular uncertainty between $2-3^{\circ}$. The opening angle between each GRB and
event pair was a greater than $2\sigma$ deviation with respect to the signal space PDF. 
Though the pre-trials significances of these coincidences were $0.0069$ and
$0.0020$ for GRB131202B and GRB150428B, respectively, correcting for trials
these are $0.988$ and $0.930$. The remaining coincident events of Tables~\ref{tab:nt_results} and
\ref{tab:st_results} are low significance coincidences, as measured by the event
signal-to-background PDF ratios and post-trials p-values. 
%These coincidences
%fall into two categories: 1)~track events in coincidence with poorly-localized
%GRBs detected by Fermi GBM, or 2)~poorly reconstructed track events in
%coincidence with a GRB. 
In summary, the set of per-GRB coincidences observed is
taken to be consistent with background. 

The only coincidence that contributes significantly to a non-zero per-year and
channel stacked test statistic is the one coincident with GRB110207A. The
significance of the combined Northern and Southern Hemisphere track stacked
analysis test statistic is $p=0.42$. Combined with the previously published
four-year Northern Hemisphere track~\citep{IC.NorthTrackGRB.2015} and three-year
all-sky cascade~\citep{IC.AllskyCascadeGRB.2016} analyses, the stacked analysis
has a final significance of $p=0.60$, consistent with the background-only
hypothesis. For the GRB sample analyzed in this paper, the benchmark standard
fireball, photospheric fireball, and ICMART models were expected to yield 2.75,
4.66, and 0.10 neutrino events, respectively. When combined with
previously published searches, these models are expected to yield 6.51, 11.02, 
and 0.25 neutrino events, respectively. Though a number of events have been found
temporally coincident with GRBs, none has appeared to be a particularly
compelling signal and they have occurred at a rate consistent with background.

Having found results consistent with background, limits can be placed on
neutrino production models in GRBs. These amount to calculating the Neyman upper
limit~\citep{Neyman.Stats.1937} on the flux normalization of these models by
determining the fraction of Monte Carlo pseudo-experiments in which such a model
would yield a test statistic at least as extreme as that observed. For example,
a model can be excluded at the 90\% confidence level (CL) should it result in
90\% of pseudo-experiments with $\Tcal \geq \Tcal_{\mathrm{obs}}$. Limits
calculated account for systematic uncertainties in the ice model, DOM
efficiency, and interaction cross sections, which translate to a
10\%--20\% uncertainty in model limits. The effect of these systematic
uncertainties in calculated model limits is determined in a model-dependent way,
as their effect is found to be much more pronounced at low energy than at high energy.

\begin{figure}[t]
  \centering
  \includegraphics[width=0.37\textwidth]{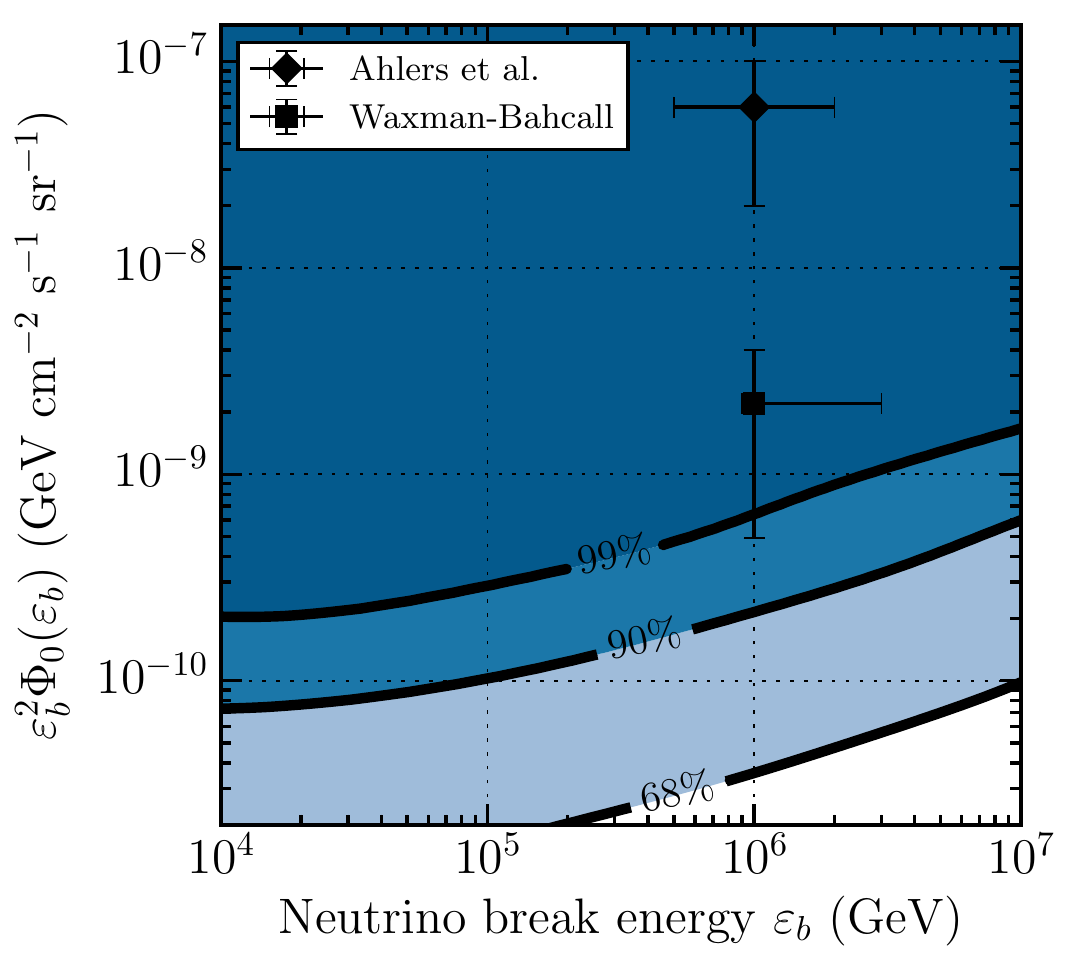}
  \caption{Excluded regions for a given CL of the generic double broken power
    law neutrino spectrum as a function of first break energy $\varepsilon_b$
    and per-flavor quasi-diffuse flux normalization $\Phi_0$ derived from the
    presented results combined with previous Northern Hemisphere
    track~\citep{IC.NorthTrackGRB.2015} and all-sky
    cascade~\citep{IC.AllskyCascadeGRB.2016} searches. Models of neutrino
    production assuming GRBs are the sole source of the measured UHECR flux
    either by neutron escape~\citep{Ahlers.GRBNu.2011} or proton
    escape~\citep{Waxman.GRBNu.1997} from the relativistic fireball are provided
    for reference.} 
  \label{fig:uhecr_model_exclusion_contours}
\end{figure}

Constraints were first determined for a generic double broken power law neutrino
flux of the form%~\citep{Waxman.GRBNu.1997,Ahlers.GRBNu.2011}

\begin{equation}
  \Phi_{\nu} (E_{\nu}) = \Phi_0 \times
  \begin{cases}
    \varepsilon_b^{-1} E_{\nu}^{-1}, & E_{\nu} \leq \varepsilon_b \\
    E_{\nu}^{-2}, & \varepsilon_b < E_{\nu} \leq 10\varepsilon_b \\
    E_{\nu}^{-4} \left( 10 \varepsilon_b \right)^2, & 10\varepsilon_b < E_{\nu},
  \end{cases}
  \label{dbpl_nu}
\end{equation}
as a function of first break energy $\varepsilon_b$ and quasi-diffuse spectral
normalization $\Phi_0$. These limits are presented in
\figref{uhecr_model_exclusion_contours} as excluded regions in this parameter
space. Two models of neutrino production in GRBs where GRBs are assumed to be
the sole origin of the measured UHECR flux are provided in this parameter space:
the neutron escape model of \cite{Ahlers.GRBNu.2011} and the proton escape model
of \cite{Waxman.GRBNu.1997}, which has been updated with recent measurements of
the UHECR flux~\citep{Katz.UHECRs.2009}. Both models are excluded at over 90\%
confidence level (CL) with most of the model assumption phase space excluded at
over the 99\% CL. A thorough reconsideration of whether GRBs can be the sources
of UHECRs from~\cite{Baerwald.UHECR.2015} shows that the internal shock fireball
model is still plausible if cosmic ray protons can efficiently escape the
fireball with a low pion-production efficiency for a range of $f_{p}$ and
$\Gamma$, which predict neutrino fluxes below the current limits.
%greatly constraining the hypothesis that GRBs are significant producers of UHECRs in the prompt phase. 

Similar constraints were calculated for simple power law spectra consistent with
IceCube's observed astrophysical neutrino flux~\citep{IC.3yrHESE.2014,
IC.CosmicMuonNu.2015, IC.MESE.2015, IC.CosmicMuonNu.2016}, concluding that
$\lesssim 0.4\%$ of the astrophysical neutrino flux can be the result of a GRB
prompt, quasi-diffuse flux assuming no spectral breaks. This constraint is
weakened to a $\lesssim 1\%$ contribution should there be a low-energy spectral
break in the astrophysical neutrino flux below $\unit[100]{TeV}$.

\begin{figure}[t]
  \centering
  \includegraphics[width=0.47\textwidth]{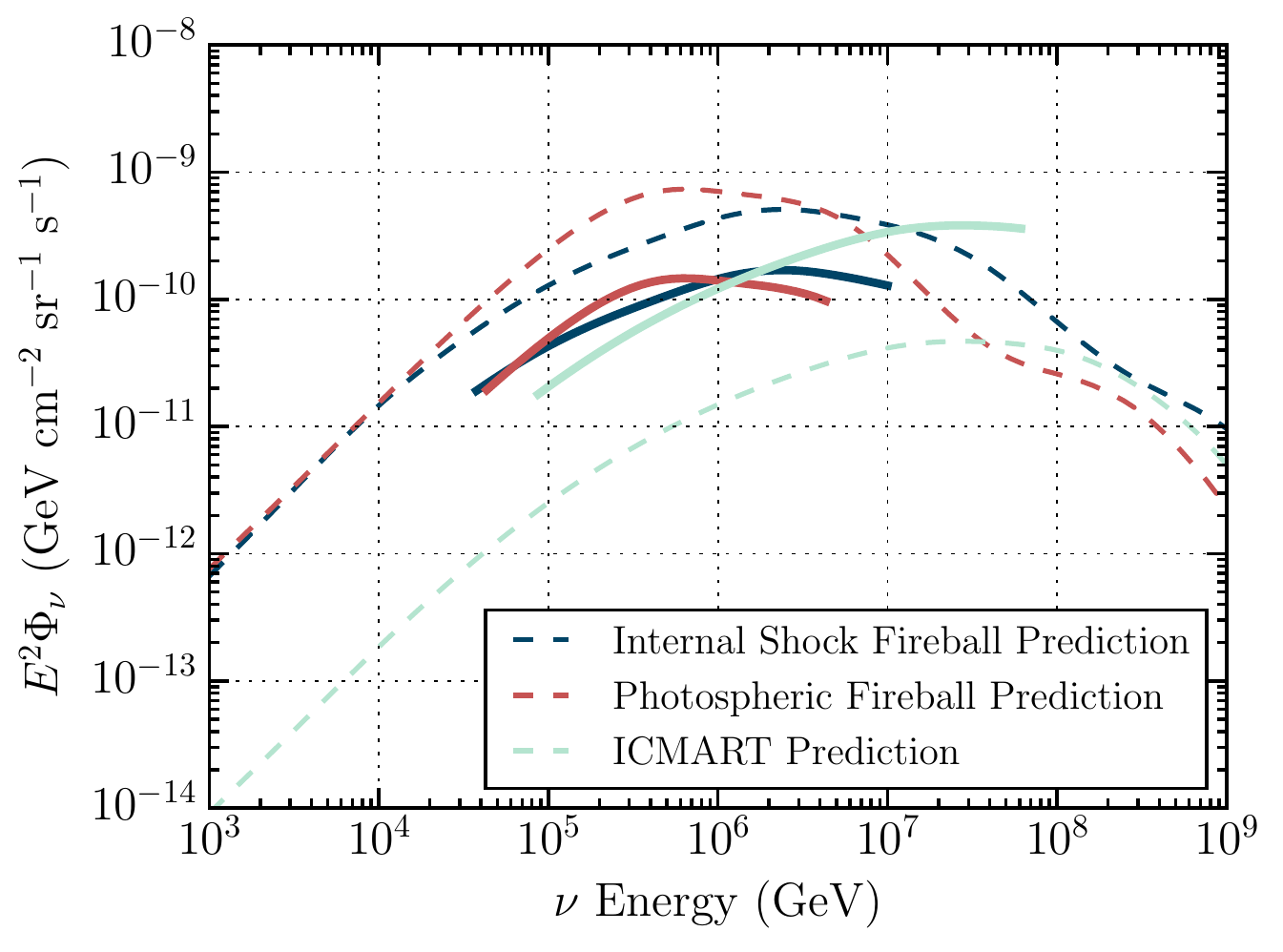}
  \caption{Upper limits (90\% CL, solid lines) to the predicted per-flavor
    quasi-diffuse flux of numerical neutrino production models (dashed lines)
    for benchmark parameters $f_p = 10$ and $\Gamma = 300$  over the expected
    central 90\% central energy containment interval of detected neutrinos for
    these models, combining the presented analysis with the previously published
    Northern Hemisphere $\nu_{\mu}$ track~\citep{IC.NorthTrackGRB.2015} and
    all-sky cascade~\citep{IC.AllskyCascadeGRB.2016} searches.} 
  \label{fig:final_model_lims}
\end{figure}

We also calculated limits for the numerical models of neutrino production in
GRBs, where the expected measurable neutrino fluence is determined from the
per-GRB $\gamma$-ray spectrum parameters. First, upper limits (90\% CL) are
calculated for the internal shock fireball, photospheric fireball, and ICMART
models using benchmark parameters of the fireball baryonic loading $f_p = 10$ and
bulk Lorentz factor $\Gamma = 300$. These are presented in
\figref{final_model_lims}, scaling the model fluences to a per-flavor
quasi-diffuse flux. Both the internal shock and photospheric fireball models are
strongly constrained. The ICMART model significantly reduces the expected
neutrino production in GRBs and remains beyond the sensitivity of the combined
analysis.

\begin{figure*}[t]
  \centering
  \includegraphics[width=0.325\textwidth]{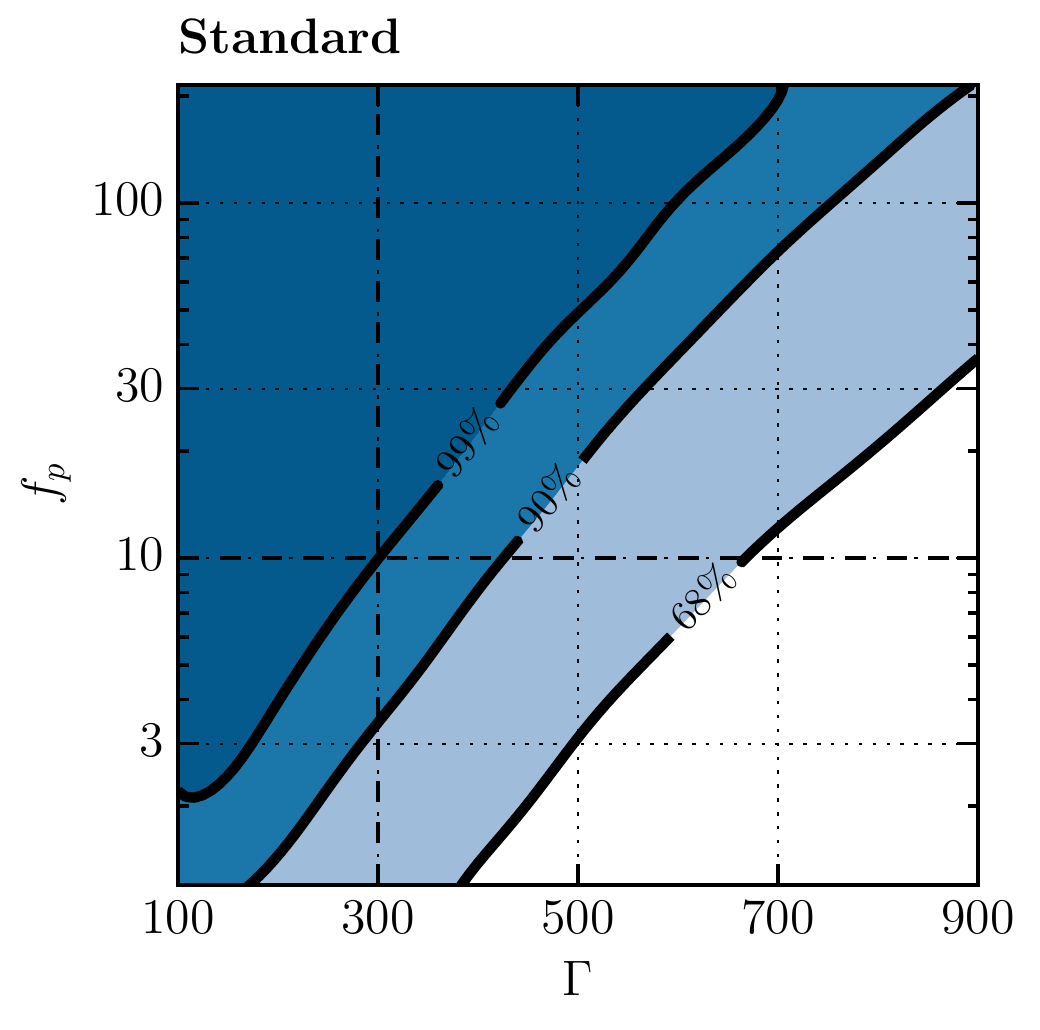}
  \includegraphics[width=0.325\textwidth]{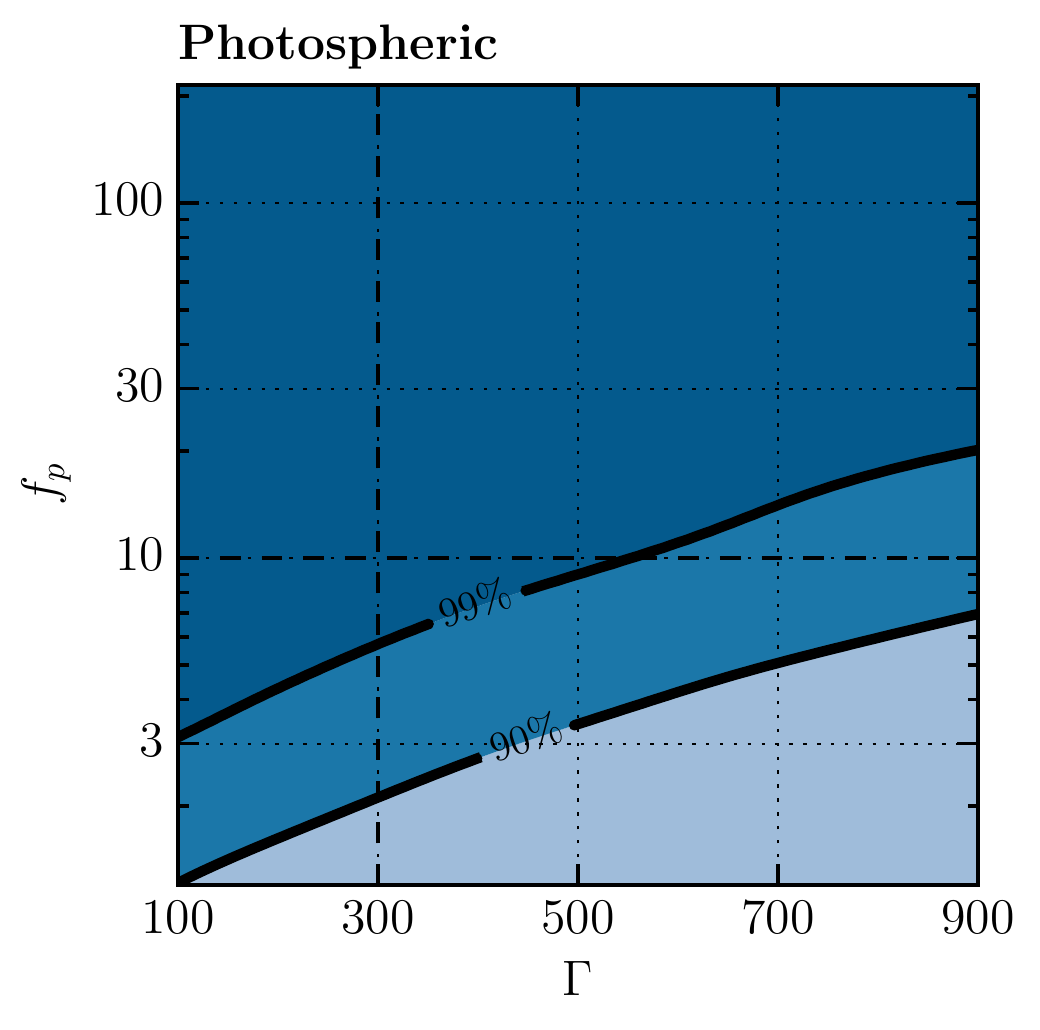}
  \includegraphics[width=0.325\textwidth]{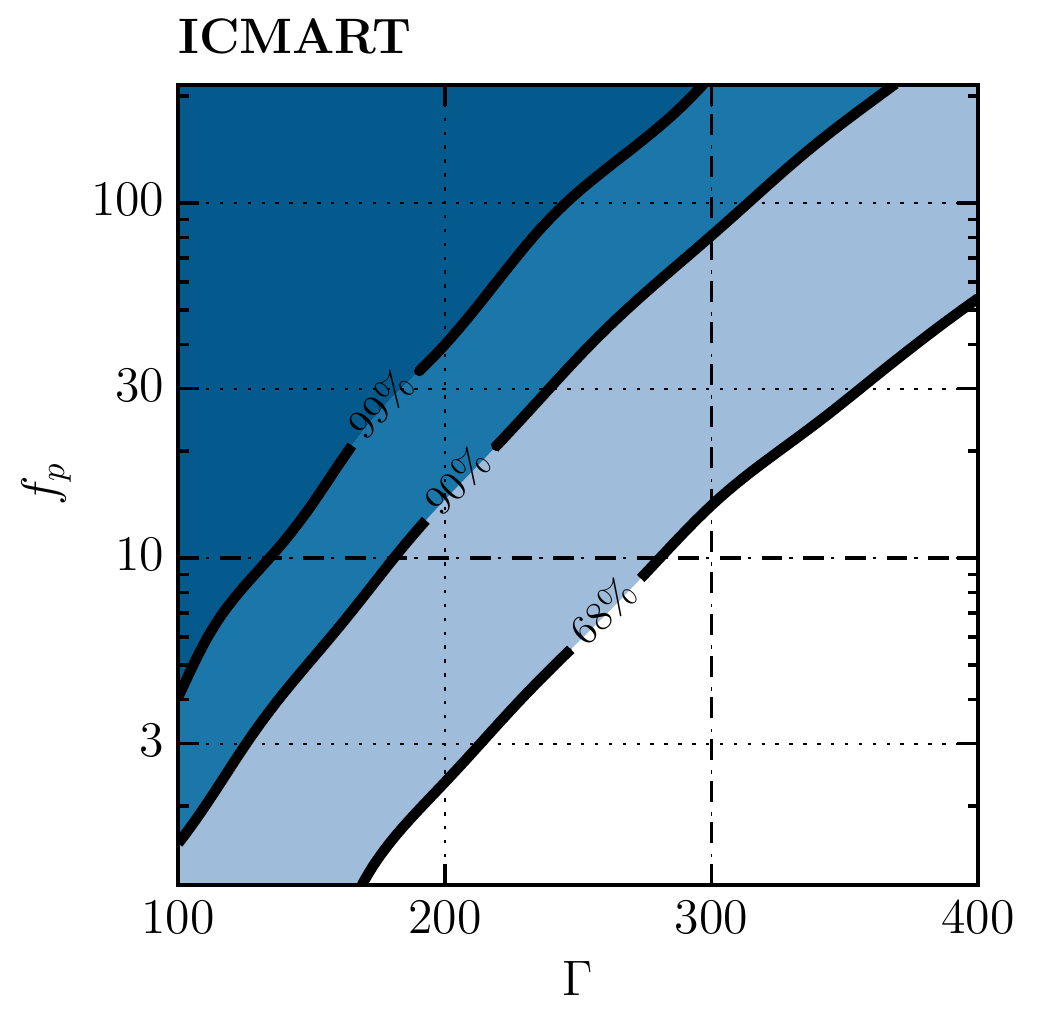}
  \caption{Excluded regions for a given CL in $f_p$ and $\Gamma$ parameter space
    for three numerical models of neutrino production in GRBs, derived from the
    presented results combined with previous Northern Hemisphere
    track~\citep{IC.NorthTrackGRB.2015} and all-sky
    cascade~\citep{IC.AllskyCascadeGRB.2016} searches. Left: internal shock
    fireball model, Middle: photospheric fireball model, and Right: ICMART model.}
  \label{fig:model_exclusion_contours}
\end{figure*}

These limits are extended to arbitrary values for $f_b$ and $\Gamma$ in the
numerical models. Assuming all GRBs in the analyzed sample have identical values
for $f_p$ and $\Gamma$, limits are presented in
\figref{model_exclusion_contours} as exclusion regions in a scan of $f_p$ and
$\Gamma$ parameter space. Here, the internal shock and photospheric fireball
models are shown to be excluded at the 99\% CL for benchmark model parameters.
The 90\% CL upper limits of all models are improved by about a factor of two compared to
those presented in the all-sky cascade analysis~\citep{IC.AllskyCascadeGRB.2016}
with the inclusion of this new three year Northern Hemisphere and five year Southern
sky $\nu_{\mu} + \bar{\nu}_{\mu}$ analysis.
%while the 90\% CL exclusion region is extended compared to those published in
%the All-sky cascade analysis~\citep{IC.AllskyCascadeGRB.2016}. 
The primary regions in these models that still cannot be constrained require
small baryonic loading and large bulk Lorentz factors. The ICMART model is
limited in a much smaller interval of possible bulk Lorentz factors ($100 <
\Gamma < 400$) as this model is much less well constrained; only regions of
large baryonic loading and small bulk Lorentz factors can be meaningfully
excluded. 

\section{Conclusions} 
\label{sec:conclusions}

We have performed a search for muon neutrinos and anti-neutrinos in coincidence
with 1172 GRBs in IceCube data. This analysis consisted of an extension of
previous Northern Hemisphere track analyses to three more years of data, and aa
additional search for $\nu_{\mu} + \bar{\nu}_{\mu}$ induced track events in the
Southern Hemisphere in five years of IceCube data, which improves the
sensitivity of the analysis to neutrinos with energy above a few PeV. Taken
together, these searches greatly improve IceCube's sensitivity to neutrinos
produced in GRBs when combined with previous analyses. A number of events were
found temporally coincident with these GRBs, but were consistent with background
both individually and when stacked together. New limits were therefore placed on
prompt neutrino production models in GRBs, which represent the strongest
constraints yet on the proposal that GRBs are the primary source of UHECRs during
their prompt phase. General models of neutrino emission were first constrained
as a function of spectral break energy and flux normalization, excluding much of
the current model phase space where GRBs during their prompt emission are
assumed to be the sole source of UHECRs in the universe at the 99\% CL. Furthermore,
models deriving an expected prompt neutrino flux from individual GRB
$\gamma$-ray spectral properties were constrained as a function of GRB outflow
hadronic content and Lorentz factor $\Gamma$. Models of prompt neutrino
production that have not yet been excluded require GRBs to have much lower
neutrino production efficiency, either through reduced hadronic content in the
outflow, increased $\Gamma$-factor, or acceleration regions much farther from
the central engine than the standard internal shock fireball model predicts.
This analysis also does not meaningfully address the possible GRB production of
neutrinos during their precursor or afterglow phases.

The continuing exclusion of the internal shock fireball model, as well as its
photospheric extension, is not altogether surprising, as the radiative
efficiency of these models has long been suggested to be insufficient to yield
the observed $\gamma$-ray spectra~\citep{FanPiran.RadEff.2006,
Swift.RadEff.2007, Kumar.GRBs.2015}, unless the distribution of shell
$\Gamma$-factors within the fireball is unrealistically
broad~\citep{Beloborodov.IS.2000, Kumar.GRBs.2015}.  Furthermore,
\cite{Baerwald.UHECR.2015} have self-consistently constrained the hypothesis
that GRBs are the source of UHECRs under the single-zone internal shock fireball
model from the measured UHECR spectrum, $\gamma$-ray measurements, and IceCube
limits to GRB neutrino production and cosmogenic neutrinos. These constraints in
addition to null results in new IceCube prompt neutrino and cosmogenic neutrino
searches increasingly require unphysically large baryonic loading factors or
fireball bulk Lorentz factors that may be in tension with multi-wavelength
measurements of $\Gamma \gtrsim 10$ for some
GRBs~\citep{Laskar.MultiWaveGRB.2015}. Multi-zone internal shock fireball models
of neutrino production remain beyond the sensitivity of this
work~\citep{Bustamante.MultiZoneNuGRB.2015, Globus.MultiZoneNuGRB.2015}, and
thus are unconstrained.

This paper has introduced a new method for analyzing GRBs on an individual
basis, which is adaptable to near real-time analyses for neutrino production in
detected GRBs. Though the analysis in this paper improved constraints on
neutrino production in GRBs, such constraints operate under the assumption of
roughly uniform production across GRBs. Should a rare subclass of GRBs produce a
significant neutrino signal, it may still be discoverable
with fast follow-up by IceCube and multi-wavelength observations. The all-sky
$\nu_{\mu} + \bar{\nu}_{\mu}$ CC interaction channel investigated in this paper
is especially promising for this purpose. In addition, the proposed IceCube
extension, IceCube-\emph{Gen2}~\citep{IC.Gen2.2014, IC.PSLoc.2014}, would
increase the detector's sensitivity to transient astrophysical neutrino sources 
and possibly reveal GRB neutrino production below IceCube's current sensitivity.
The continued non-detection of a prompt neutrino signal, however, will increasingly
disfavor GRBs as a source of UHECRs.
% TODO: re-write last sentence (taken from hellauer)

\input{acknowledgments}

\pagebreak

\bibliographystyle{apj}
\bibliography{grb}

\begin{deluxetable*}{cccccc}
  \tablecaption{Three-year Northern Hemisphere track analysis coincident
    events. The duration $T_{100}$, angular uncertainty $\sigma$ (Fermi-GBM
    statistical-only uncertainties indicated by $*$), and total $\gamma$-ray fluence
    ($\unit[]{erg\, cm^{-2}}$) of GRBs
    with coincident events are described. Coincident events are summarized in
    terms of their time relative to the GRB start time $T_1$, their angular
    uncertainty $\sigma$, angular displacement from the GRB location $\Delta
    \Psi$, and reconstructed muon energy proxy (TeV). Event significance is estimated by
    their signal-to-background PDF ratio value $\Scal / \Bcal$ (only events with $\Scal / \Bcal > 10^{-3}$ are listed), while final GRB
    coincidence significance is given as pre-trials (post-trials) p-values
    relative to background-only test statistic distributions. \label{tab:nt_results}}
  \tablehead{
    & \colhead{Time} & \colhead{$\sigma$} & \colhead{$\Delta \Psi$} &
    \colhead{Fluence/Energy} & \colhead{Significance}
  }
  \startdata
    GRB120612B & $T_{100} = \unit[63.24]{s}$ & ${}^*$$7.1^{\circ}$ & &
    $2.06 \times 10^{-6}$ & $p = 0.049$ $(1)$ \\
    Event 1 & $T_1 + \unit[47.71]{s}$ & $5.3^{\circ}$ & $29.0^{\circ}$ &
    $\gtrsim 0.54$ & $\Scal/\Bcal = 1.4$ \\
    \hline
    GRB120911A & $T_{100} = \unit[28.58]{s}$ & $0.0003^{\circ}$ & &
    $2.34 \times 10^{-6}$ & $p = 0.0044$ $(1)$ \\
    Event 1 & $T_1 + \unit[120.94]{s}$ & $4.6^{\circ}$ & $2.9^{\circ}$ &
    $\gtrsim 0.98$ & $\Scal/\Bcal = 3.1$ \\
    \hline
    GRB130116A & $T_{100} = \unit[66.82]{s}$ & ${}^*$$29.9^{\circ}$ & &
    $9.27 \times 10^{-7}$ & $p = 0.076$ $(1)$ \\
    Event 1 & $T_1 + \unit[69.25]{s}$ & $0.5^{\circ}$ & $67.7^{\circ}$ &
    $\gtrsim 2.1$ & $\Scal/\Bcal = 1.5$ \\
    \hline
    GRB130318A & $T_{100} = \unit[137.99]{s}$ & ${}^*$$9.9^{\circ}$ & &
    $3.41 \times 10^{-6}$ & $p = 0.021$ $(1)$ \\
    Event 1 & $T_1 + \unit[29.83]{s}$ & $0.6^{\circ}$ & $18.5^{\circ}$ &
    $\gtrsim 0.46$ & $\Scal/\Bcal = 6.4$ \\
    Event 2 & $T_1 + \unit[44.58]{s}$ & $2.5^{\circ}$ & $48.2^{\circ}$ &
    $\gtrsim 0.32$ & $\Scal/\Bcal = 0.024$ \\
    \hline
    GRB130925B & $T_{100} = \unit[265.47]{s}$ & ${}^*$$4.1^{\circ}$ & &
    $1.49 \times 10^{-5}$ & $p = 0.032$ $(1)$ \\
    Event 1 & $T_1 + \unit[108.8]{s}$ & $3.4^{\circ}$ & $12.6^{\circ}$ &
    $\gtrsim 0.70$ & $\Scal/\Bcal = 16.3$ \\
    \hline
    GRB131029B & $T_{100} = \unit[50.95]{s}$ & ${}^*$$5.8^{\circ}$ & &
    $4.49 \times 10^{-6}$ & $p = 0.053$ $(1)$ \\
    Event 1 & $T_1 + \unit[50.49]{s}$ & $2.4^{\circ}$ & $18.2^{\circ}$ &
    $\gtrsim 0.68$ & $\Scal/\Bcal = 4.9$ \\
    \hline
    GRB131202B & $T_{100} = \unit[86.02]{s}$ & ${}^*$$2.2^{\circ}$ & &
    $1.24 \times 10^{-5}$ & $p = 0.0069$ $(0.988)$ \\
    Event 1 & $T_1 + \unit[85.18]{s}$ & $2.1^{\circ}$ & $7.5^{\circ}$ &
    $\gtrsim 1.7$ & $\Scal/\Bcal = 122.1$ \\
    %Event 2 & $T_1 - \unit[81.68]{s}$ & $1.2^{\circ}$ & $73.1^{\circ}$ &
    %$\gtrsim 1.4$ & $\Scal/\Bcal = 6.9 \times 10^{-10}$ \\
    \hline
    GRB140404B & $T_{100} = \unit[26.63]{s}$ & ${}^*$$2.2^{\circ}$ & &
    $8.18 \times 10^{-6}$ & $p = 0.026$ $(1)$ \\
    Event 1 & $T_1 - \unit[38.49]{s}$ & $5.4^{\circ}$ & $13.1^{\circ}$ &
    $\gtrsim 1.1$ & $\Scal/\Bcal = 11.0$ \\
    %Event 2 & $T_1 + \unit[92.08]{s}$ & $0.8^{\circ}$ & $67.8^{\circ}$ &
    %$\gtrsim 0.40$ & $\Scal/\Bcal = 5.7 \times 10^{-9}$ \\
    \hline
    GRB140521B & $T_{100} = \unit[46.59]{s}$ & ${}^*$$10.1^{\circ}$ & &
    $2.75 \times 10^{-6}$ & $p = 0.051$ $(1)$ \\
    Event 1 & $T_1 + \unit[98.37]{s}$ & $1.6^{\circ}$ & $11.5^{\circ}$ &
    $\gtrsim 0.79$ & $\Scal/\Bcal = 7.3$ \\
    \hline
    GRB140603A & $T_{100} = \unit[138.24]{s}$ & ${}^*$$2.1^{\circ}$ & &
    $1.86 \times 10^{-5}$ & $p = 0.025$ $(1)$ \\
    Event 1 & $T_1 + \unit[41.35]{s}$ & $1.1^{\circ}$ & $14.9^{\circ}$ &
    $\gtrsim 1.5$ & $\Scal/\Bcal = 10.1$ \\
    Event 2 & $T_1 - \unit[33.78]{s}$ & $1.3^{\circ}$ & $38.7^{\circ}$ &
    $\gtrsim 2.1$ & $\Scal/\Bcal = 0.026$ \\
    %Event 3 & $T_1 + \unit[28.93]{s}$ & $0.6^{\circ}$ & $54.7^{\circ}$ &
    %$\gtrsim 0.68$ & $\Scal/\Bcal = 5.5 \times 10^{-5}$ \\
    \hline
    GRB141029B & $T_{100} = \unit[202.44]{s}$ & ${}^*$$1.0^{\circ}$ & &
    $3.8 \times 10^{-5}$ & $p = 0.034$ $(1)$ \\
    Event 1 & $T_1 - \unit[10.33]{s}$ & $1.6^{\circ}$ & $11.7^{\circ}$ &
    $\gtrsim 0.70$ & $\Scal/\Bcal = 6.4$ \\
    Event 2 & $T_1 - \unit[80.99]{s}$ & $1.0^{\circ}$ & $30.2^{\circ}$ &
    $\gtrsim 0.45$ & $\Scal/\Bcal = 0.003$ \\
    \hline
    GRB150428B & $T_{100} = \unit[161.8]{s}$ & $0.0003^{\circ}$ & &
    $3.7 \times 10^{-6}$ & $p = 0.0020$ $(0.930)$ \\
    Event 1 & $T_1 + \unit[71.35]{s}$ & $2.9^{\circ}$ & $6.0^{\circ}$ &
    $\gtrsim 3.2$ & $\Scal/\Bcal = 131.9$ \\
    \hline
    GRB150428D & $T_{100} = \unit[32.51]{s}$ & ${}^*$$6.1^{\circ}$ & &
    $1.53 \times 10^{-6}$ & $p = 0.024$ $(1)$ \\
    Event 1 & $T_1 - \unit[43.69]{s}$ & $4.4^{\circ}$ & $15.2^{\circ}$ &
    $\gtrsim 0.54$ & $\Scal/\Bcal = 9.4$ \\
    \hline
    GRB150507A & $T_{100} = \unit[63.49]{s}$ & ${}^*$$1.4^{\circ}$ & &
    $1.52 \times 10^{-5}$ & $p = 0.039$ $(1)$ \\
    Event 1 & $T_1 + \unit[58.24]{s}$ & $1.1^{\circ}$ & $20.4^{\circ}$ &
    $\gtrsim 1.8$ & $\Scal/\Bcal = 2.4$ \\
    Event 2 & $T_1 - \unit[74.44]{s}$ & $2.1^{\circ}$ & $36.3^{\circ}$ &
    $\gtrsim 0.69$ & $\Scal/\Bcal = 0.0023$ \\
    %Event 3 & $T_1 - \unit[12.1]{s}$ & $3.5^{\circ}$ & $66.5^{\circ}$ &
    %$\gtrsim 2.2$ & $\Scal/\Bcal = 3.1 \times 10^{-6}$ \\
    %Event 4 & $T_1 + \unit[4.6]{s}$ & $1.6^{\circ}$ & $67.3^{\circ}$ &
    %$\gtrsim 0.84$ & $\Scal/\Bcal = 3.4 \times 10^{-7}$
  \enddata
\end{deluxetable*}

\begin{deluxetable*}{cccccc}
  \tablecaption{Southern Hemisphere track analysis ontime events, following
  the conventions of \tabref{nt_results}. \label{tab:st_results}}
  \tablehead{
    & \colhead{Time} & \colhead{$\sigma$} & \colhead{$\Delta \Psi$} &
    \colhead{Fluence/Energy} & \colhead{Significance}
  }
  \startdata
    GRB110105A & $T_{100} = \unit[123.39]{s}$ & ${}^*$$2.0^{\circ}$ & &
    $2.09 \times 10^{-5}$ & $p = 0.037$ $(1)$ \\
    Event 1 & $T_1 + \unit[102.0]{s}$ & $0.3^{\circ}$ & $13.1^{\circ}$ &
    $\gtrsim 15$ & $\Scal/\Bcal = 2.2$ \\
    \hline
    GRB110207A & $T_{100} = \unit[109.32]{s}$ & $0.0132^{\circ}$ & &
    $4.4 \times 10^{-6}$ & $p = 0.00035$ $(0.540)$ \\
    Event 1 & $T_1 + \unit[87.4]{s}$ & $0.3^{\circ}$ & $0.9^{\circ}$ &
    $\gtrsim 12$ & $\Scal/\Bcal = 271.6$ \\
    \hline
    GRB111205A & $T_{100} = \unit[80.38]{s}$ & $0.1^{\circ}$ & &
    $1.7 \times 10^{-4}$ & $p = 0.0023$ $(1)$\\
    Event 1 & $T_1 + \unit[150.9]{s}$ & $18.7^{\circ}$ & $17.3^{\circ}$ &
    $\gtrsim 482$ & $\Scal/\Bcal = 9.5$ \\
    \hline
    GRB121127A & $T_{100} = \unit[3.51]{s}$ & $0.08^{\circ}$ & &
    $9.34 \times 10^{-7}$ & $p = 0.00043$ $(1)$ \\
    Event 1 & $T_1 + \unit[2.42]{s}$ & $60.1^{\circ}$ & $79.5^{\circ}$ &
    $\gtrsim 175$ & $\Scal/\Bcal = 0.85$ \\
    \hline
    GRB121231A & $T_{100} = \unit[32.77]{s}$ & ${}^*$$6.5^{\circ}$ & &
    $2.94 \times 10^{-6}$ & $p = 0.035$ $(1)$ \\
    Event 1 & $T_1 + \unit[66.5]{s}$ & $0.5^{\circ}$ & $13.9^{\circ}$ &
    $\gtrsim 24$ & $\Scal/\Bcal = 4.2$ \\
    \hline
    GRB130909A & $T_{100} = \unit[33.79]{s}$ & ${}^*$$17.2^{\circ}$ & &
    $1.98 \times 10^{-6}$ & $p = 0.010$ $(0.989)$ \\
    Event 1 & $T_1 + \unit[14.9]{s}$ & $0.2^{\circ}$ & $19.4^{\circ}$ &
    $\gtrsim 53$ & $\Scal/\Bcal = 30.6$ \\
    \hline
    GRB130924A & $T_{100} = \unit[37.1]{s}$ & ${}^*$$6.0^{\circ}$ & &
    $3.73 \times 10^{-6}$ & $p = 0.033$ $(1)$ \\
    Event 1 & $T_1 + \unit[92.6]{s}$ & $27.1^{\circ}$ & $8.0^{\circ}$ &
    $\gtrsim 72$ & $\Scal/\Bcal = 1.3$ \\
    Event 2 & $T_1 + \unit[6.6]{s}$ & $0.4^{\circ}$ & $19.3^{\circ}$ &
    $\gtrsim 2.8$ & $\Scal/\Bcal = 0.84$ \\
    \hline
    GRB131119A & $T_{100} = \unit[34.8]{s}$ & ${}^*$$7.3^{\circ}$ & &
    $1.85 \times 10^{-6}$ & $p = 0.025$ $(1)$ \\
    Event 1 & $T_1 - \unit[23.1]{s}$ & $0.4^{\circ}$ & $22.9^{\circ}$ &
    $\gtrsim 16$ & $\Scal/\Bcal = 8.2$ \\
    \hline
    GRB141012A & $T_{100} = \unit[37.64]{s}$ & ${}^*$$3.1^{\circ}$ & &
    $6.64 \times 10^{-6}$ & $p = 0.014$ $(1)$ \\
    Event 1 & $T_1 + \unit[100.54]{s}$ & $11.5^{\circ}$ & $22.4^{\circ}$ &
    $\gtrsim 114$ & $\Scal/\Bcal = 2.5$ \\
    \hline
    GRB141013A & $T_{100} = \unit[82.43]{s}$ & ${}^*$$3.8^{\circ}$ & &
    $8.81 \times 10^{-6}$ & $p = 0.017$ $(1)$ \\
    Event 1 & $T_1 + \unit[34.4]{s}$ & $17.6^{\circ}$ & $48.0^{\circ}$ &
    $\gtrsim 459$ & $\Scal/\Bcal = 2.3$ \\
    %Event 2 & $T_1 + \unit[97.9]{s}$ & $1.2^{\circ}$ & $75.9^{\circ}$ &
    %$\gtrsim 1.0$ & $\Scal/\Bcal = 2.1 \times 10^{-9}$ \\
    \hline
    GRB150222C & $T_{100} = \unit[74.75]{s}$ & ${}^*$$11.32^{\circ}$ & &
    $3.84 \times 10^{-6}$ & $p = 0.020$ $(1)$ \\
    Event 1 & $T_1 + \unit[22.73]{s}$ & $0.3^{\circ}$ & $24.5^{\circ}$ &
    $\gtrsim 31$ & $\Scal/\Bcal = 8.4$ \\
    Event 2 & $T_1 - \unit[61.2]{s}$ & $0.2^{\circ}$ & $52.3^{\circ}$ &
    $\gtrsim 50$ & $\Scal/\Bcal = 0.0064$
  \enddata
\end{deluxetable*}

\end{document}

%% file: authors.tex
% authors.tex

\AuthorCallLimit=1
\fullcollaborationName{IceCube}

\author{
IceCube Collaboration:
M.~G.~Aartsen\altaffilmark{1},
M.~Ackermann\altaffilmark{2},
J.~Adams\altaffilmark{3},
J.~A.~Aguilar\altaffilmark{4},
M.~Ahlers\altaffilmark{5},
M.~Ahrens\altaffilmark{6},
I.~Al~Samarai\altaffilmark{7},
D.~Altmann\altaffilmark{8},
K.~Andeen\altaffilmark{9},
T.~Anderson\altaffilmark{10}
% ...
}
\altaffiltext{1}{Department of Physics, University of Adelaide, Adelaide, 5005, Australia}
\altaffiltext{2}{DESY, D-15735 Zeuthen, Germany}
\altaffiltext{3}{Dept.~of Physics and Astronomy, University of Canterbury, Private Bag 4800, Christchurch, New Zealand}
\altaffiltext{4}{Universit\'e Libre de Bruxelles, Science Faculty CP230, B-1050 Brussels, Belgium}
\altaffiltext{5}{Dept.~of Physics and Wisconsin IceCube Particle Astrophysics Center, University of Wisconsin, Madison, WI 53706, USA}
\altaffiltext{6}{Oskar Klein Centre and Dept.~of Physics, Stockholm University, SE-10691 Stockholm, Sweden}
\altaffiltext{7}{D\'epartement de physique nucl\'eaire et corpusculaire, Universit\'e de Gen\`eve, CH-1211 Gen\`eve, Switzerland}
\altaffiltext{8}{Erlangen Centre for Astroparticle Physics, Friedrich-Alexander-Universit\"at Erlangen-N\"urnberg, D-91058 Erlangen, Germany}
\altaffiltext{9}{Department of Physics, Marquette University, Milwaukee, WI, 53201, USA}
\altaffiltext{10}{Dept.~of Physics, Pennsylvania State University, University Park, PA 16802, USA}
% ...

%% file: acknowledgments.tex
\acknowledgments

We acknowledge the support from the following agencies:
U.S. National Science Foundation-Office of Polar Programs,
U.S. National Science Foundation-Physics Division,
University of Wisconsin Alumni Research Foundation,
the Grid Laboratory Of Wisconsin (GLOW) grid infrastructure at the University of Wisconsin - Madison, the Open Science Grid (OSG) grid infrastructure;
U.S. Department of Energy, and National Energy Research Scientific Computing Center,
the Louisiana Optical Network Initiative (LONI) grid computing resources;
Natural Sciences and Engineering Research Council of Canada,
WestGrid and Compute/Calcul Canada;
Swedish Research Council,
Swedish Polar Research Secretariat,
Swedish National Infrastructure for Computing (SNIC),
and Knut and Alice Wallenberg Foundation, Sweden;
German Ministry for Education and Research (BMBF),
Deutsche Forschungsgemeinschaft (DFG),
Helmholtz Alliance for Astroparticle Physics (HAP),
Research Department of Plasmas with Complex Interactions (Bochum), Germany;
Fund for Scientific Research (FNRS-FWO),
FWO Odysseus programme,
Flanders Institute to encourage scientific and technological research in industry (IWT),
Belgian Federal Science Policy Office (Belspo);
University of Oxford, United Kingdom;
Marsden Fund, New Zealand;
Australian Research Council;
Japan Society for Promotion of Science (JSPS);
the Swiss National Science Foundation (SNSF), Switzerland;
National Research Foundation of Korea (NRF);
Villum Fonden, Danish National Research Foundation (DNRF), Denmark